\documentstyle[12pt]{article}

\def\hybrid{\topmargin -20pt    \oddsidemargin 0pt
        \headheight 0pt \headsep 0pt
        \textwidth 6.25in       
        \textheight 9.5in       
        \marginparwidth .875in
        \parskip 5pt plus 1pt   \jot = 1.5ex}

\hybrid

\def\baselinestretch{1.2}

\catcode`\@=11

\def\marginnote#1{}
\newcount\hour
\newcount\minute
\newtoks\amorpm
\hour=\time\divide\hour by60
\minute=\time{\multiply\hour by60 \global\advance\minute by-\hour}
\edef\standardtime{{\ifnum\hour<12 \global\amorpm={am}%
        \else\global\amorpm={pm}\advance\hour by-12 \fi
        \ifnum\hour=0 \hour=12 \fi
        \number\hour:\ifnum\minute<10 0\fi\number\minute\the\amorpm}}
\edef\militarytime{\number\hour:\ifnum\minute<10 0\fi\number\minute}

\def\draftlabel#1{{\@bsphack\if@filesw {\let\thepage\relax
   \xdef\@gtempa{\write\@auxout{\string
      \newlabel{#1}{{\@currentlabel}{\thepage}}}}}\@gtempa
   \if@nobreak \ifvmode\nobreak\fi\fi\fi\@esphack}
        \gdef\@eqnlabel{#1}}
\def\@eqnlabel{}
\def\@vacuum{}
\def\draftmarginnote#1{\marginpar{\raggedright\scriptsize\tt#1}}

\def\draft{\oddsidemargin -.5truein
        \def\@oddfoot{\sl preliminary draft \hfil
        \rm\thepage\hfil\sl\today\quad\militarytime}
        \let\@evenfoot\@oddfoot \overfullrule 3pt
        \let\label=\draftlabel
        \let\marginnote=\draftmarginnote
   \def\@eqnnum{(\theequation)\rlap{\kern\marginparsep\tt\@eqnlabel}%
\global\let\@eqnlabel\@vacuum}  }

\def\preprint{\twocolumn\sloppy\flushbottom\parindent 2em
        \leftmargini 2em\leftmarginv .5em\leftmarginvi .5em
        \oddsidemargin -.5in    \evensidemargin -.5in
        \columnsep .4in \footheight 0pt
        \textwidth 10.in        \topmargin  -.4in
        \headheight 12pt \topskip .4in
        \textheight 6.9in \footskip 0pt
        \def\@oddhead{\thepage\hfil\addtocounter{page}{1}\thepage}
        \let\@evenhead\@oddhead \def\@oddfoot{} \def\@evenfoot{} }

\def\numberbysection{\@addtoreset{equation}{section}
        \def\theequation{\thesection.\arabic{equation}}}

\def\underline#1{\relax\ifmmode\@@underline#1\else
        $\@@underline{\hbox{#1}}$\relax\fi}

\def\titlepage{\@restonecolfalse\if@twocolumn\@restonecoltrue\onecolumn
     \else \newpage \fi \thispagestyle{empty}\c@page\z@
        \def\thefootnote{\fnsymbol{footnote}} }

\def\endtitlepage{\if@restonecol\twocolumn \else \newpage \fi
        \def\thefootnote{\arabic{footnote}}
        \setcounter{footnote}{0}}  

\catcode`@=12
\relax

\def\figcap{\section*{Figure Captions\markboth
        {FIGURECAPTIONS}{FIGURECAPTIONS}}\list
        {Figure \arabic{enumi}:\hfill}{\settowidth\labelwidth{Figure
999:}
        \leftmargin\labelwidth
        \advance\leftmargin\labelsep\usecounter{enumi}}}
 \relax
\def\tablecap{\section*{Table Captions\markboth
        {TABLECAPTIONS}{TABLECAPTIONS}}\list
        {Table \arabic{enumi}:\hfill}{\settowidth\labelwidth{Table
999:}
        \leftmargin\labelwidth
        \advance\leftmargin\labelsep\usecounter{enumi}}}
 \relax
\def\reflist{\section*{References\markboth
        {REFLIST}{REFLIST}}\list
        {[\arabic{enumi}]\hfill}{\settowidth\labelwidth{[999]}
        \leftmargin\labelwidth
        \advance\leftmargin\labelsep\usecounter{enumi}}}
 \relax
\makeatletter
\newcounter{pubctr}
\def\publist{\@ifnextchar[{\@publist}{\@@publist}}
\def\@publist[#1]{\list
        {[\arabic{pubctr}]\hfill}{\settowidth\labelwidth{[999]}
        \leftmargin\labelwidth
        \advance\leftmargin\labelsep
        \@nmbrlisttrue\def\@listctr{pubctr}
        \setcounter{pubctr}{#1}\addtocounter{pubctr}{-1}}}
\def\@@publist{\list
        {[\arabic{pubctr}]\hfill}{\settowidth\labelwidth{[999]}
        \leftmargin\labelwidth
        \advance\leftmargin\labelsep
        \@nmbrlisttrue\def\@listctr{pubctr}}}
 \relax
\makeatother
\newskip\humongous \humongous=0pt plus 1000pt minus 1000pt

\newif\ifdtup

\relax

\def\be{\begin{equation}}
\def\ee{\end{equation}}
\def\ba{\begin{eqnarray}}
\def\ea{\end{eqnarray}}

\def\no{\noindent}

\def\IR{\relax{\rm I\kern-.18em R}}

\def\IR{\relax{\rm I\kern-.18em R}}
\def\inv{^{\raise.15ex\hbox{${\scriptscriptstyle -}$}\kern-.05em 1}}

\begin{document}

\renewcommand{\theequation}{\thesection.\arabic{equation}}

\newcommand{\beq}{\begin{equation}}
\newcommand{\eeq}[1]{\label{#1}\end{equation}}
\newcommand{\ber}{\begin{eqnarray}}
\newcommand{\eer}[1]{\label{#1}\end{eqnarray}}
\newcommand{\eqn}[1]{(\ref{#1})}
\begin{titlepage}
\begin{center}

\hfill hep--th/0507284\\
\hfill CERN-PH-TH/2005-134\\
\hfill July 2005\\

\vskip .4in

{\large \bf The algebraic structure of geometric flows in two dimensions}

\vskip 0.6in

{\bf Ioannis Bakas}\footnote{On sabbatical leave from Department of Physics, 
University of Patras, GR-26500 Patras, Greece; e-mail: 
bakas@ajax.physics.upatras.gr}
\vskip 0.2in
{\em Theory Division, Department of Physics, CERN \\
CH-1211 Geneva 23, Switzerland\\
\footnotesize{\tt ioannis.bakas@cern.ch}}\\

\end{center}

\vskip .8in

\centerline{\bf Abstract}

\no
There is a common description of different intrinsic geometric flows in two  
dimensions using Toda field equations associated to continual Lie algebras that  
incorporate the deformation variable $t$ into their system. The Ricci flow admits    
zero curvature formulation in terms of an infinite dimensional algebra with Cartan  
operator $\partial/\partial t$. Likewise, the Calabi flow arises as Toda field   
equation associated to a supercontinual algebra with odd Cartan  
operator $\partial/\partial \theta - \theta \partial/\partial t$. Thus, taking the  
square root of the Cartan operator allows to connect the two distinct classes of 
geometric deformations of second and fourth order, respectively. The algebra is also   
used to construct formal solutions of the Calabi flow in terms of free fields by 
B\"acklund transformations, as for the Ricci flow. Some applications of the present 
framework to the general class of Robinson-Trautman metrics that describe spherical  
gravitational radiation in vacuum in four space-time dimensions are also discussed. 
Further iteration of the algorithm allows to construct an infinite hierarchy of 
higher order geometric flows, which are integrable in two dimensions and they admit 
immediate generalization to K\"ahler manifolds in all dimensions. These flows 
provide examples of more general deformations introduced  by Calabi that preserve   
the K\"ahler class and minimize the quadratic curvature functional for extremal 
metrics.  
\vfill
\end{titlepage}
\eject

\def\baselinestretch{1.2}
\baselineskip 16 pt
\noindent


\section{Introduction}
\setcounter{equation}{0}

The subject of geometric flows witnessed rapid development in recent years. It  
only fair to say that many areas of mathematics underwent rapid development as 
new tools became available for 
addressing long standing open problems - predominantly in differential geometry - 
via appropriately chosen evolution equations, also known as geometric flows. 
At the same time, several 
cases of such flows arose independently in physics with applications 
ranging from classical mechanics to general relativity and quantum field theory.
Roughly speaking, there are two general classes of deformation equations in 
differential geometry, namely intrinsic and extrinsic curvature flows. The former     
refer to geometric evolutions of Riemannian metrics $g$ on a given manifold $M$ 
that are driven by the Ricci curvature, in various forms, whereas the latter 
refer to geometric evolutions of submanifolds embedded in $M$ that are driven 
by their extrinsic curvature. As such, they all correspond to dynamical  
systems in {\em superspace}, which is the infinite dimensional space of all 
possible metrics on a given manifold. 
The geometric flows provide systems of parabolic equations, being first order in 
the deformation variable $t$ that is typically called time, and they are second 
or higher order in the space variables. 
However, the equations are quite 
difficult to solve in all generality, due to non-linearities. Although the  
short time existence of solutions for a given initial metric is guaranteed  
by the parabolic nature of the equations, 
their convergence to canonical metrics after sufficiently long time has 
only been analyzed under various 
conditions, and in connection to the problem of formation of singularities 
along the flows. 

All geometric flows share some common qualitative 
features with the linear heat flow equation, which has the tendency to dissipate 
any temperature perturbations all over space after sufficiently long time.    
Thus, curvature perturbations around canonical metrics tend to wash away after 
infinitely long time, provided that no singularities are formed along the way 
and that the corresponding geometric deformations are taken in some normalized form, so 
that the volume of the space remains invariant in time. It is precisely for this 
reason that continuous flows to constant curvature metrics have been particularly
successful for exploring the relations among the geometric and topological structures 
of manifolds and prove various geometrization conjectures in low dimensions, 
as in the classic uniformization problem of Riemann surfaces. However, there is  
no general dictionary at this moment that connects geometric evolution 
equations to physical problems, apart from some special examples that are 
under intense investigation in recent years. 
It is quite plausible that field theory, classical as well as quantum, may 
provide a unifying framework for all different species of geometric flows by giving 
appropriate physical interpretation to the deformation variable $t$ and to the 
structure of their driving terms. 

The Ricci flow is the prime example of this kind as it describes 
the beta function equations of two dimensional sigma models to lowest order in perturbation 
theory. The Calabi flow, on the other hand, makes its appearance in general relativity while 
studying the general class of four-dimensional Robinson-Trautman radiative metrics. 
Both these flows correspond to intrinsic geometric deformations and they will 
occupy most of the present work. In fact, it proves advantageous to treat 
both of them in parallel,  
as we uncover some formal relations among the two, which 
exist in two dimensions and generalize 
quite naturally to K\"ahler manifolds in all dimensions. Thus, some known 
techniques for the integration of Ricci flows will be found to 
admit immediate generalization 
to the Calabi flow, using the appropriate mathematical framework that will be 
developed in due course. 
As byproduct, the formal construction of solutions  
describing spherical gravitational waves in vacuum will be achieved  
in the context of four-dimensional general relativity, thus providing the main physical 
application of this work.    
Although our analysis is only applicable, as it stands, to intrinsic 
geometric flows, it is worth noting in passing that other important classes 
of evolution equations, such as those correspond to extrinsic curvature flows, 
might become tractable by similar methods.   
Such generalizations are mostly interesting for a variety of problems in boundary 
quantum field theory, but they will not be included here. 
 
The primary aim is to develop some new algebraic techniques for casting 
intrinsic geometric flows in two dimensions into zero curvature form 
and prove their integrability, in a 
certain sense, using infinite dimensional Lie algebras. 
These results should 
be considered complementary to the traditional techniques used in the mathematics literature 
for studying general properties and qualitative features of geometric evolution equations.  
Although strictly limited to two dimensions, 
where most of the algebraic techniques are becoming  
available, the results might also be of more general value 
given the embedding of two-dimensional manifolds into higher dimensional
spaces that deform geometrically in one way or another. 
The present description views the geometric flows as  
Toda field equations associated to infinite dimensional algebras that incorporate the 
deformation variable $t$ into the system of their defining commutation relations. 
This approach works well for the two-dimensional Ricci flow, but it also generalizes 
quite naturally to the Calabi flow using a formal relation between the two via 
super-evolution. In fact, as will be seen later, one follows from the other by taking the 
square root of the time evolution operator, and, thus, their formal integration 
by group theoretical methods appears to have many similarities. This relation will be 
further generalized to induce an infinite hierarchy of higher order evolution 
equations and associated infinite dimensional Lie algebras that enable to cast them 
in zero curvature form. 

The algebras used for the purposes of the present work deserve further study as they fall
in the general (but poorly understood) class of infinite dimensional Lie algebras with  
infinite growth. Two-dimensional integrable systems based on infinite dimensional 
Lie algebras are quite interesting to study on their own, in all generality, and in 
comparison to the better understood cases of integrable systems that admit zero 
curvature formulation with finite dimensional 
gauge algebras. A key point in our analysis is 
the uneven treatment of the space coordinates and the time $t$, which is fully 
encoded into the structure of the infinite dimensional algebra used for the 
effective  
two-dimensional description of the equations. Thus, the systems appear to be 
integrable in the two-dimensional sense, whereas at the same time exhibit 
dissipative behavior in time without reaching a contradiction of terms.    
Finally, the algebraic structures underlying these flows might be of more 
general value if could be extended to encompass other interesting dynamical phenomena in 
geometrical theories with relevance to the physical world, such as gravitation. 

Actually, the framework that is developed here, based on Toda field equations for 
infinite dimensional algebras, admits a smooth limit when $t$-independent solutions 
of the two-dimensional Calabi flow are investigated. This 
class of solutions is also of paramount 
importance in general relativity as they describe Robinson-Trautman space-times 
of a certain algebraic type. In this case, as will be seen in detail, there is a zero 
curvature formulation of the equation using Kac's $K_2$ simple Lie algebra with  
infinite growth. The results that we derive will also be compared to other works on this 
subject and they will be further used to construct formal power series solutions in 
terms of free fields as in ordinary Toda systems. As byproduct, a new description of 
the $K_2$ algebra will be obtained within the general framework of supercontinual 
Lie algebras that are introduced here for the first time and used for the zero curvature 
formulation of the Calabi flow, its variants, and descendants. Thus, the systematic   
study and classification of such novel algebraic structures seems to be a 
valuable problem of common interest in physics and mathematics.    

Some generalizations to geometric flows on K\"ahler manifolds
of arbitrary dimension will also be considered. In that general context, the hierarchy of 
geometric flows provide special examples of a more general deformation problem posed 
by Calabi for minimizing the quadratic curvature functional, while preserving the 
K\"ahler class of the metrics. The standard Calabi flow is the most elementary 
example in this class of deformations with known physical applications. All of them, 
however, are naturally associated to Calabi's variational problem in the space of all 
possible metrics on a given K\"ahler manifold with Euler-Lagrange equations 
satisfied by extremal metrics, such as metrics with constant curvature. Thus, they 
have equal importance in geometry.  

The structure of this paper is organized as follows. In section 2, after  
briefly reviewing the general set up of the main equations, the emphasis 
is subsequently placed 
on deformations of two-dimensional geometries, where the new algebraic techniques are
becoming available for their (formal) integration by group theoretical methods.   
In section 3, we give account of the notion of continual Lie algebras, their main properties 
and the systems of Toda field equations associated to them in zero curvature form. 
The general method of integration is also illustrated for general choices of Cartan operators
and further generalizations to the class of supercontinual Lie algebras are introduced.    
In section 4, applications of the algebraic method to the two-dimensional Ricci flow are  
briefly summarized together with the formal construction of the general solution in terms
of free fields by B\"acklund transformation. In section 5, 
the method is extended to the two-dimensional Calabi flow by 
introducing an infinite dimensional algebra whose structure is dictated by the super-evolution 
operator that connects it to the Ricci flow. In this context, it is more natural to view  
the zero curvature formulation of the Calabi flow as providing two-dimensional integrable 
system for a supercontinual Lie algebra,
where the generators depend on the coordinates $(t, \theta)$ of $R^{1|1}$  
superspace. This reformulation also proves useful for the description of the general 
solution in terms of free fields, as for the Ricci flow. In section 6, the relation of 
the two-dimensional Calabi flow to the theory of spherical gravitational waves in vacuum 
is spelled out in detail using the class of Robinson-Trautman metrics that represent 
gravitational radiation in the exterior of bounded sources in four space-time dimensions. 
Then, genuine solutions of the Calabi flow correspond to 
type II gravitational backgrounds in Petrov's classification of space-time geometries. 
Type III metrics, which correspond to time independent solutions of the Calabi flow, 
are studied separately in section 7 by Toda field theory techniques and the connections 
to Kac's $K_2$ algebra are spelled out in great detail. 
In section 8, an infinite hierarchy of higher order geometric flows is constructed in 
two dimensions by iterating the formal relation between the Ricci and Calabi flows,  
via super-evolutions, and their zero curvature formulation is studied in all 
generality. In section 9, this hierarchy is shown to admit immediate generalization 
to K\"ahler manifolds in all dimensions and the resulting evolution equations 
provide special examples of a more general problem posed by Calabi for deforming  
arbitrary K\"ahler metrics toward extremal metrics.  
Finally, section 10 contains our conclusions and a brief outline of some directions 
for further work.

\section{Intrinsic geometric evolutions}
\setcounter{equation}{0}

In this section we present an account of the Ricci and Calabi flows, as they are 
defined in K\"ahler geometry, and then reveal a formal relation between the two 
that plays important role later. Although our results are mostly confined to 
one complex (or two real) dimensions, a more general, but brief, introduction is necessary 
in order to understand which of them can be extended to arbitrary dimensions.    

\subsection{Preliminaries}

Let $M$ denote a complex $n$-dimensional manifold, which for all practical purposes is 
taken to be connected and compact without boundary. Assume also that $M$ admits a 
K\"ahler metric $g$, which is locally expressible in the form
\be
ds^2 = 2 g_{a\bar{b}} dz^a d\bar{z}^b ~, 
\ee
using a system of holomorphic coordinates $z^a$ and their complex conjugates $\bar{z}^a$ 
with $a=1,2, \cdots, n$. The K\"ahler condition on the metric is expressed by the existence
of a locally defined, real valued, smooth function $K(z, \bar{z})$, which is not unique, such 
that $g_{a \bar{b}} (z, \bar{z}) = \partial_a \bar{\partial}_b K (z, \bar{z})$. We also 
consider the associated real valued exterior form of degree $(1,1)$, 
\be
\omega = \sqrt{-1} ~ g_{a\bar{b}} dz^a \wedge d \bar{z}^b ~,  
\ee
which is closed and determines the principal cohomology class of the metric, $[\omega]$.
As such, it is an intrinsic invariant of the complex analytic structure on $M$.  
We are interested only in the space of K\"ahler metric in a fixed cohomology class; 
other K\"ahler metrics of the same class can be obtained by the transformation 
\be
\tilde{g}_{a\bar{b}} = g_{a\bar{b}} + {\partial^2 u \over \partial z^a 
\partial \bar{z}^b} ~, \label{trneme}  
\ee
where $u$ is a globally defined real valued function on $M$.   

For later use it is convenient to recall the expression for the Ricci curvature 
tensor, in complex notation,  
\be
R_{a \bar{b}} = - {\partial^2 {\rm log} ({\rm det}g) \over \partial z^a 
\partial \bar{z}^b}    
\ee
and the Ricci scalar curvature, which is 
\be
R = {1 \over 2} g^{\mu \nu} R_{\mu \nu} = 
g^{a\bar{b}} R_{a\bar{b}} ~. 
\ee
Note that the normalization used here is one half of the usual value found in most texts.  
Finally, the volume element is 
\be
dV(g) = \omega^{[n]} = (\sqrt{-1})^n {\rm det}(g_{a \bar{b}}) dz^1 \wedge d\bar{z}^1 \wedge 
\cdots \wedge dz^n \wedge d\bar{z}^n   
\ee
and it is also an invariant of the complex analytic structure on $M$.  

The Laplace-Beltrami operator for a K\"ahler metric $g$ on $M$ is defined as usual, using 
the analyst's sign,  
\be
\Delta \varphi = {1 \over 2} g^{ij} \nabla_i \nabla_j \varphi = 
g^{a\bar{b}} \nabla_a \bar{\nabla}_b \varphi = 
g^{a\bar{b}} {\partial^2 \varphi \over \partial z^a \partial \bar{z}^b} ~.  
\ee
The integral $\int \Delta \varphi ~ dV(g)$ over the entire space vanishes 
by Stoke's theorem, since it has been assumed that $M$ is compact without boundaries.
Also, for later use, it is convenient to 
recall the definition of the fourth order operator, $D_L$,
which is an elliptic differential operator that acts on scalar functions as
follows, 
\ba
D_L \varphi & = & g^{a\bar{n}} {\partial \over \partial z^a} \left(g^{b \bar{c}} 
{\partial \over \partial z^b}  
\left(g_{l \bar{n}} {\partial \over \partial \bar{z}^c}  (g^{l \bar{m}} 
{\partial \varphi \over \partial \bar{z}^m} ) \right) \right) \nonumber\\
& = & 
\Delta \Delta \varphi + \varphi^{, a \bar{b}} R_{a \bar{b}} + \varphi^{, a} R_{,a} ~.  
\ea
It is a strongly elliptic operator that coincides with its complex conjugate $\bar{D}_L$
if and only if $R$ is constant and, furthermore, satisfies the relation 
$D_L R = \bar{D}_L R$ for any K\"ahler metric.      

The geometric flows that will be considered next provide deformations of the K\"ahler 
metrics within a given cohomology class $[\omega]$. As such they can be formulated 
as parabolic evolution equations for the single function $u(z, \bar{z}; t)$ 
that arises in the class of 
transformations \eqn{trneme}, or equivalently 
$\tilde{\omega} = \omega + \partial \bar{\partial} 
u$. However, we will follow the standard presentation based on evolution 
equations for the components of the metric itself, which are driven by the intrinsic 
curvature of $M$ in various forms, depending on the type of flow,  
and which characterize the order of the non-linear differential equations.  

In all cases, the metric is prescribed at some initial time, say $t=0$, and evolves 
accordingly to later times. Short time solutions always exist, due to the parabolic 
nature of the equations, but their long time existence is not guaranteed and 
depends on various conditions and on the dimensionality of the manifold $M$. When they
exist, however, have the tendency to converge to canonical metrics defined in 
appropriate ways. The advantage 
of using K\"ahler geometry is that all information about the curvature is encoded 
into a single function, the scalar curvature $R$, as in two dimensions. However, 
many important details also depend on the dimensionality of the K\"ahler manifold
and its properties.

\subsection{Ricci flow} 
The Ricci flow is the simplest example of a geometric deformation, which is driven by 
the Ricci curvature tensor 
associated to a metric $g$ on the manifold $M$ and assumes the following form 
(see, for instance, \cite{ricci, chow1}, and references therein)   
\be
\partial_t g_{a\bar{b}} = - R_{a\bar{b}} ~.   
\ee
Actually, this equation can be defined on any Riemannian manifold of arbitrary 
dimension, but we only consider here its application to K\"ahler geometry. 
More general, the Ricci flow takes the form  
\be
\partial_t g_{\mu \nu} = - R_{\mu \nu} + \nabla_{\mu} \xi_{\nu} + 
\nabla_{\nu} \xi_{\mu} ~, \label{modrfe}  
\ee
thus also taking into account the effect of 
arbitrary (time dependent) reparametrizations associated to a vector field 
$\xi$. It is a parabolic second order non-linear differential equation  
that arose independently in physics and mathematics for different reasons. 

In physics it has definite meaning within the renormalization group analysis of  
two-dimensional sigma models, as it provides the rate of change of their target space 
metric - viewed as generalized coupling - with respect to the logarithm of the 
world-sheet length scale, \cite{polyakov, friedan}. 
The Ricci curvature tensor arises as the beta function 
of the target space metric, to lowest order in perturbation theory of the 
quantum theory, and drives the evolution of the generalized coupling 
from the ultra-violet to the infra-red regime. 
Solutions to this equation do not correspond to conformal field theories, but they
rather describe transitions between different closed string vacua in the weak  
gravitational regime of the theory induced by tachyon condensation. In this context,
reparametrizations in target space that are associated to gradient vector fields,
$\xi_{\mu} = \nabla_{\mu} \phi$, can be interpreted as having a dilaton field $\phi$ 
in target space coupled to the metric. Likewise, fixed points of the generalized 
Ricci flow equation \eqn{modrfe}, satisfying the Ricci soliton equation 
$R_{\mu \nu} = 2 \nabla_{\mu} \nabla_{\nu} \phi$, correspond to non-trivial 
conformal field theories. The Euclidean black-hole in two-dimensional target
space with the shape of a semi-infinite long cigar provides the simplest example of 
such a Ricci-K\"ahler soliton with axial symmetry.    

Another interesting application arises from the possibility to embed the Ricci flow 
into Einstein equations using 
two extra dimensions with coordinates $u$ and $v$, as follows, 
\be
ds^2 = -2 du dv + g_{\mu \nu} (x; u) dx^{\mu} dx^{\nu} ~,  
\ee
provided that the light-cone coordinate $u$ is identified with $t$ and there is also 
a dilaton field linear in $v$ coupled to the higher dimensional metric, \cite{tseyt}. 
This prescription allows to construct gravitational backgrounds - though of special 
type - for consistent string propagation, using solutions of the Ricci flow equation
in two dimensions lower. Then, the renormalization group flow becomes the profile of 
a gravitational wave in space-time, and the geometry interpolates between fixed 
points of the flow in transverse space as one traverses the wave. 
In effect, this construction makes the Ricci flow to appear as special case of 
Einstein equations, with the appropriate ansatz, and it is the closest  
one can get to a similar derivation of the (two-dimensional) Calabi flow from Einstein 
equations, as will be seen later.    

In mathematics, it is often appropriate to use the normalized Ricci flow, which
is defined as follows, \cite{ricci, chow1},
\be
\partial_t g_{\mu \nu} = - R_{\mu \nu} + {<R> \over {\rm dim} M} g_{\mu \nu} ~,    
\label{norifl}
\ee
where $<R>$ is the average (mean) scalar curvature
\be
<R> = {\int_M R[g] ~ dV(g) \over \int_M dV(g)} ~,  
\ee
which is independent of the 
coordinates on the manifold and can only depend on $t$.  
The compensating term is designed so that the volume of the space $M$ is preserved 
under continuous deformations, unlike the standard (unnormalized) Ricci flow that 
changes it in time. Thus, the normalized flow is more useful as tool for proving 
various uniformization theorems in geometry, since its solutions do not become 
extinct after some finite time, but they have better chance to exist for sufficiently 
long time and converge to canonical (constant curvature) metrics that correspond 
to the fixed points of equation \eqn{norifl} on $M$. However, the two variants of 
the flow are related to each other by performing  
rescaling of the metric and time reparametrization according to the rule
\be
\tilde{g}_{\mu \nu} = f(t) g_{\mu \nu} ~, ~~~~~ \tilde{t} = \int dt f(t) ~, 
\ee
which sets $<\tilde{R}>(\tilde{t}) = ({\rm dim} M) f^{\prime}(t)/f^2(t)$. 
    
In K\"ahler geometry, in particular, the Ricci flow has been employed for studying
the existence of K\"ahler-Einstein metric on manifolds under appropriate technical
conditions. It has already been applied successfully to two dimensional Riemann surfaces
of genus $0$, $1$ or $g \geq 2$ by re-deriving the classic uniformization theorem 
of Poincar\'e that asserts the existence of constant curvature metrics with values 
$+1$, $0$ or $-1$, respectively; as a result, these surfaces are quotients of 
$S^2$, $R^2$ or $H^2$ by a discrete subgroup $\Gamma$ acting freely and isometrically.
In higher dimensions the situation is promising but more complex.

\subsection{Calabi flow}

The Calabi flow, on the other hand, which is only defined for K\"ahler manifolds,  
assumes the following form, \cite{calabi1, calabi2},  
\be
\partial_t g_{a \bar{b}} = {\partial^2 R \over 
\partial z^a \partial \bar{z}^b} ~. \label{calaflow}  
\ee
It is volume preserving deformation without need for adjustment, unlike the Ricci flow, 
and it also preserves the K\"ahler class of the metric. 
The Calabi flow is a parabolic equation for 
the components of the metric, or equivalently for the K\"ahler potential, but it is    
fourth order in the variables 
$z$ and $\bar{z}$ and, thus, more complicated to analyze in general.  
Critical points of the flow are called {\em extremal metrics} and they clearly 
encompass constant curvature metrics, if they exist on a given K\"ahler manifold. 
In this respect, the Calabi flow is used as a tool for studying the conditions  
for Einstein-K\"ahler metrics in geometry, and in conjunction with 
their possible obstructions.  

There is a variational problem in the space of K\"ahler metrics  
that allows to formulate the problem in general terms. In particular, consider 
the quadratic curvature functional 
\be
S (g) = \int_M R^2[g] ~ dV(g) \label{quacurfu}  
\ee
defined for all possible metrics on $M$ with fixed cohomology class $[\omega]$. By 
computing the variation of $S (g)$ with respect to the flow \eqn{calaflow} one finds
\be
\partial_t S (g) = \int_M \left(- 2R \Delta \Delta R - 2 R R^{, a\bar{b}} R_{a \bar{b}} 
+ R^2 \Delta R \right) dV(g) ~,  
\ee
since the associated scalar curvature flows as   
\be
\partial_t R = - \Delta \Delta R - R^{, a\bar{b}} R_{a \bar{b}} ~.  
\ee
Next, it is useful to employ the fourth order operator $D_L$ to cast the result of 
the variation into the form 
\be
\partial_t S (g) = \int_M \left(- 2R D_L R + 2 R R^{, a} R_{, a} 
+ R^2 \Delta R \right) dV(g)  
\ee
and observe that $(R^2 R^{, a})_{, a} = 2R R^{, a} R_{, a} + R^2 \Delta R$. Thus, 
dropping the total derivative term, it follows immediately that
\be
\partial_t S (g) = -2 \int_M R D_L R ~ dV(g) ~.   
\ee

The operator $D_L$ is positive semi-definite and self-adjoint operator acting 
on functions on $M$. It can be further decomposed as  
$D_L = L^{\star} L$, where $L$ is a second order operator mapping complex valued functions 
to holomorphic tangent vector valued $(0,1)$-forms, 
\be
L \varphi = \varphi^{, ac} g_{c \bar{b}} {\partial \over \partial z^a} \otimes d \bar{z}^b ~, 
\ee
and $L^{\star}$ denotes its adjoint. Then, using for simplicity the Hermitian product
\be
(\varphi, ~ \psi) = \int_M \varphi (z, \bar{z}) \bar{\psi} (z, \bar{z}) ~ dV(g) ~,  
\ee
we have in total 
\be
\partial_t S (g) = \partial_t (R, ~ R) = -2 (R, ~ D_LR) = -2 (LR, ~ LR) \leq 0 ~,      
\ee
showing that $S (g)$ is a Lyapunov functional that decreases monotonically along 
the Calabi flow. 

The critical points correspond to extremal metrics on $M$, \cite{calabi1, calabi2}, 
which according 
to the previous discussion satisfy the Euler-Lagrange equation $D_L R = 0$, 
or equivalently
the simpler equation $LR = 0$ that describes its global solutions in a compact 
manifold $M$ without boundary. This, in turn implies that the extremal metrics,  
when they exist, are characterized 
by the condition that their curvature is such that $R^{, a} \partial_a$ is a holomorphic 
vector field. In general, very little is known about the space of extremal 
K\"ahler metrics, but it is expected to have unique extremal metric in each K\"ahler 
class up to holomorphic transformations, apart from some special examples.   
Further analysis of the problem shows that if there exists a constant scalar curvature 
metric on $M$, then it achieves the absolute minimum value of $S (g)$. In this case,
all extremal metrics have constant curvature and the absolute minimum value of the 
quadratic curvature functional that follows by application of Schwarz's 
inequality,   
\be
S(g) \geq {\left(\int_M R[g] ~ dV(g)\right)^2 \over \int_M dV(g)}  ~,  
\ee
is attained. However, there can be situations where 
extremal metrics exist on $M$ but they do not have constant scalar curvature; in fact, 
constant scalar curvature metrics and extremal metrics with non-constant scalar 
curvature do not co-exist in a single K\"ahler class. 

Whenever the existence of a critical K\"ahler metric can be guaranteed, then the so called 
Futaki-Calabi obstructions determine the necessary and sufficient 
conditions for the existence of a constant scalar curvature metric in K\"ahler geometry, 
\cite{futaki, calabi2}.  
Last, but not 
least, there are counter-examples of K\"ahler manifolds that do not admit any extremal 
metrics and, hence, $S (g)$ has no critical points; for, if extremal metrics exist 
they will necessarily be symmetric under a maximal compact subgroup of the holomorphic 
transformation group, but, in some cases, this fails to exist, \cite{levine}. 
Given these facts, it appears 
that the situation is rather complex in general and we will not attempt to fill in the  
technical details as they are beyond the scope of the present work; for an overview, see,  
for instance, \cite{fut, tian}. Here, we will be 
mostly concerned with two-dimensional spaces with the topology of sphere, which 
admit a constant curvature metric - that of the round sphere - and permit the solutions 
of the Calabi flow to exist for sufficiently long time and converge to the canonical 
metric, \cite{chrus, chen}. 
The conditions for global existence and convergence of the Calabi flow to 
constant curvature metrics is a hard problem, in general, in higher dimensional spaces.     

The physical relevance of the Calabi flow is not yet known for manifolds with  
arbitrary number of dimensions apart from the lowest dimensional case of  
one complex (or two real) dimensions. It is quite remarkable that the two-dimensional 
Calabi flow characterizes completely the solutions of four dimensional Einstein equations
that describe gravitational radiation from bounded sources, using the general class
of Robinson-Trautman metrics for space-times that admit a geodesic, shear and twist free
diverging null congruence. In this context, $t$ is identified with the retarded time 
coordinate in space-time, as will be explained in detail in section 6.

\subsection{A formal relation between the two flows}

There is a curious relation between the Ricci and Calabi flows on K\"ahler manifolds 
of arbitrary dimension that manifests by squaring the time evolution operator. In 
particular, taking the time derivative of the Ricci flow, we have 
\be
{\partial^2 \over \partial t^2} g_{a\bar{b}} = -{\partial \over \partial t} R_{a\bar{b}} 
= \partial_a \bar{\partial}_b {\partial \over \partial t} \left({\rm log ~ det} g \right) ~,  
\ee
which in turn can be evaluated by noting  
$\partial_t ({\rm log} ~ {\rm det}g) = g^{a\bar{b}} \partial_t g_{a\bar{b}} = 
- g^{a\bar{b}} R_{a\bar{b}} = - R$. 
Thus, in K\"ahler geometry, deformations of the metric flowing \`a la Ricci always imply 
the equivalent relation 
\be
{\partial^2 \over \partial t^2} g_{a\bar{b}} = - \partial_a \bar{\partial}_b R ~.  
\ee
Clearly, if the second derivative of the metric with respect to the Ricci time is 
identified with minus its first derivative with respect to the Calabi time, 
\be
{\partial^2 \over \partial t_{{\rm R}}^2} = - 
{\partial \over \partial t_{{\rm C}}} ~, 
\ee
the two flows will be formally the same. 

The procedure above looks similar in vain to the way
of extracting the square root of the Schr\"odinger equation in supersymmetric quantum 
mechanics for a free particle (for a recent account see, for instance, \cite{manton}, 
and references therein). In the latter case, this is practically achieved by
considering a super-evolution equation 
\be
{\cal D} \Psi = {\cal Q} \Psi ~, 
\ee
where ${\cal D}$ is the supertime 
derivative with the defining property ${\cal D}^2 = -i \partial / \partial t$ 
and ${\cal Q}$ is the supercharge of the theory satisfying the relation 
${\cal D} {\cal Q} + {\cal Q} {\cal D} = 0$. Then, the super-evolution equation 
provides a consistent square root of the Schr\"odinger equation, because it implies
\be
-i{\partial \Psi \over \partial t} = {\cal D}^2 \Psi = {\cal D} {\cal Q} \Psi 
= -{\cal Q} {\cal D} \Psi = -{\cal Q}^2 \Psi = -H \Psi ~,  
\ee
where $H = {\cal Q}^2$ is the Hamiltonian of the system. Of course, the order of the
equations that govern geometric evolutions is different, and there are also non-linearities
that make the problem at hand to be more complicated mathematically.  

The analogy will be made more precise later by introducing anti-commuting parameters
into the theory of geometric flows which promote time derivatives to 
super-evolution operators. This procedure, which works well in two dimensions,   
has far reaching consequences, as it will allow to cast the Calabi flow  
into zero curvature form and deduce its properties from those of
the Ricci flow. It is for this reason that both flows are discussed in parallel  
while uncovering their algebraic structures in two dimensions. 
In the process, we will also be able to construct a hierarchy of higher order 
geometric evolution equations, together with a hierarchy of new infinite 
dimensional Lie algebras, which will be used for their zero curvature formulation.    
The general method works in the same way for higher dimensional K\"ahler manifolds,   
in that there is a squaring pattern for the time derivative operator 
that enables to connect consecutive flows, as above, but there is no zero 
curvature formulation for them above two dimensions.  
Thus, most of our work will be subsequently confined to two-dimensional spaces 
and postpone until the very end the derivation of some more general results  
that are valid in all dimensions.

\subsection{Landing in two dimensions}
        
Next, specializing to two dimensions and using a system of conformally flat (K\"ahler) 
coordinates, 
\be
ds_{\rm t}^2 = 2 e^{\Phi(z, \bar{z}; t)} dz d\bar{z} ~, 
\ee
we find that the only non-vanishing component of the Ricci curvature tensor is 
$R_{z\bar{z}} = -\partial \bar{\partial} \Phi$. 
Then, the two different geometric deformations take the 
following neat form for the conformal factor of the metric, $\Phi(z, \bar{z}; t)$, respectively,   
\ba
~~~ {\rm Ricci ~~ flow}: ~~~~~ & &  \partial_t \Phi = \Delta \Phi ~, \\
{\rm Calabi ~~ flow}: ~~~~~ & & \partial_t \Phi = - \Delta \Delta \Phi ~. 
\ea
The symbol $\Delta$ stands for the Laplace-Beltrami operator, with the 
analyst's sign,  
\be
\Delta = e^{-\Phi} \partial \bar{\partial} ~, \label{lapbel}  
\ee
which depends on $\Phi$ and introduces non-linearities into the system. These equations 
can be be defined on any Riemann surface of genus $g$, but for all practical 
purposes we can concentrate to spaces with spherical topology, $S^2$. 

The dissipative character of these equations can be inferred from their linearization 
in the weak field approximation, where the curvature perturbations are assumed to be 
small. The reference metric around which such analysis should be carried, in 
order to make good physical and 
mathematical sense, depends on the flow. For the Ricci flow, it is appropriate to 
consider small fluctuations around the flat two-dimensional plane 
characterized by small and slowly varying values of $\Phi$. This region corresponds 
to $R_{\mu \nu} \simeq 0$, that is to the asymptotically free limit of a non-linear
sigma model with target metric $g_{\mu \nu}$. In the language of renormalization 
group flows, it corresponds to the ultra-violet region when the curvature of the space
is positive and to the infra-red region when the space is negatively curved. 
Then, since $\Phi \simeq 0$, the Ricci flow equation can be approximated by the linear
heat equation $\partial_t \Phi = \partial \bar{\partial} \Phi$ in 
two-dimensional flat space, whose dissipative properties are well known.

On the other hand, for the Calabi flow, it is more appropriate to consider the round
sphere $S^2$ as reference frame for introducing small curvature perturbations around it.
The choice is essentially uniquely and 
dictated by the limiting extremal metric of the two-dimensional
Calabi flow and, therefore, one is led to examine fluctuations of the metric with  
$g = (1 + f)g_0$, where $g_0$ corresponds to the line element 
$ds^2 = 2 R_0^2 dz d \bar{z} / (1 + z \bar{z})^2$ of a sphere of radius $R_0$ and 
$f$ is taken small. Let us consider, without great loss of generality, axially  
symmetric deformations of the round sphere, which are parametrized by Legendre 
polynomials $P_l(\xi)$ in the system of spherical coordinate $(\theta, \phi)$,      
\be
ds_{\rm t}^2 = R_0^2 [1 + \epsilon_l (t) P_l ({\rm cos} \theta) ] \left(d\theta^2 + 
{\rm sin}^2 \theta d\phi^2 \right) ~, \label{pemetranz}  
\ee
with corresponding small parameters $\epsilon_l (t)$. Such configurations qualify as  
approximate solutions to the Calabi flow equation, as they preserve the volume 
to linear order in $\epsilon_l$, 
\be
V_{\epsilon} = \int dV(g_{\epsilon}) = 4\pi R_0^2 + 2\pi R_0^2 \epsilon_l (t) 
(-1)^l \int_{-1}^{+1} d\xi ~ P_l (\xi) = 4\pi R_0^2  
\ee
for all $l \geq 1$. In fact, the cases $l=0$ and $l=1$ will be omitted in the 
calculations, as they both represent round spheres. Thus, we assume that $l \geq 2$.   

Then, to linear order in the 
perturbation parameters, the Calabi flow yields the following evolution  
\be
\epsilon_l (t) = \epsilon_l (0) {\rm exp} \left(-{t \over 4 R_0^4} l(l^2 -1)(l+2) 
\right) .  
\ee
The result, which approximates well the asymptotic behavior of 
the full non-linear evolution equation, 
when $t \rightarrow +\infty$, shows that all perturbations are damped exponentially 
fast and the configuration settles down to that of a round sphere. More 
general deformations corresponding to spherical harmonics $Y_l^m (\theta, \phi)$ with 
$m \neq 0$ can also be considered, but the results are essentially unaltered. All  
linearized solutions admit interesting physical interpretation in the context of 
multi-pole gravitational radiation from bounded sources, where they were first 
derived in the literature, as will be seen later in section 6. 

For comparison, a similar analysis is performed for the normalized Ricci flow, which
is volume preserving and in two dimensions assumes the form
\be
\partial_t \Phi = \Delta \Phi + {1 \over R_0^2} 
\ee
when the space has spherical topology and its volume is taken equal to $4\pi R_0^2$. 
In this case, one considers 
linearized perturbations about the round sphere of radius $R_0$, which constitutes 
the equilibrium state of the normalized flow. 
Using the same ansatz \eqn{pemetranz} as before,
it turns out that the corresponding parameters evolve, to linear order, as 
\be
\epsilon_l (t) = \epsilon_l (0) {\rm exp} \left(-{t \over 2 R_0^2} (l-1)(l+2) 
\right) ,  
\ee
and therefore, as $t \rightarrow +\infty$, the perturbations are also damped exponentially 
fast\footnote{Of course, the time of the normalized 
flow depends logarithmically on the time of the 
unnormalized flow, as $-R_0^2 {\rm log}(-t)$, according to 
the general relation between the two; thus,
a spherical configuration that starts from very large size in 
the ultra-violet region $t \rightarrow -\infty$
and diminishes linearly to zero size at $t=0$ by the Ricci flow, will correspond to a 
volume preserving deformation of the sphere under the normalized Ricci flow that 
now runs from $-\infty$ to $+\infty$ and tends to become round exponentially fast.}. 
Comparison with the Calabi flow shows the same qualitative behavior, although in this
case perturbations of a given angular momentum $l$ diminish at a slower pace; the 
dependence of $\epsilon_l(t)$ on $l$, being 
quadratic or quartic form, is a reflection of the
corresponding evolution equations, being second and fourth order, respectively.   

It is natural to expect that the dissipative properties of the linearized 
flows will be also shared by the full non-linear parabolic equations of second and 
fourth order, respectively, as they do. The difficulty with the non-linearities is 
that they prevent the construction of explicit solutions, apart from a few  
isolated simple examples. In most cases, it is quite difficult to introduce ansatz 
that effectively reduce the non-linear equations to a simpler system of 
evolution equations for a finite number of moduli. The Ricci flow appears to be 
more tractable by such methods, since there is already a good number of simple 
(yet interesting) exact solutions. The situation is more primitive for 
the Calabi flow as only a handful of exact solutions is known.

\section{Algebraic framework}
\setcounter{equation}{0}

Our primary aim in the following is to device a suitable algebraic method for 
establishing the integrability
of the two different geometric deformations in two dimensions.  
It proves useful to sketch first the algebraic techniques
that have already been used successfully for the formal integration of the Ricci flow  
and then extend their domain of applicability to the Calabi flow and further beyond.     

The notion of continual Lie algebras provides the main framework for the 
systematic formulation of the geometric flows as Toda field equations in two  
dimensions. The key idea is to treat the deformation variable $t$ 
as a continuous ``Dynkin index" of an appropriately chosen infinite dimensional
algebra and rewrite the Ricci and Calabi flows as zero curvature conditions 
in the two-dimensional system of conformally flat coordinates. Here, we present the basic 
definitions of continual Lie algebras and the Toda systems 
associated to them. We also review the general algorithm for obtaining 
their general solutions in terms of two dimensional
free fields, using group theoretical techniques for the B\"acklund transformations. 
The specific choice  
of continual algebras that enable to cast different geometric flows into zero curvature 
form will be determined in the following sections together with the formal 
power series solutions of the equations.    

\subsection{Continual Lie algebras}

The basic theory of continual Lie algebras is formulated by introducing a system 
of Cartan-Weyl generators $\{H(t), X^{\pm} (t) \}$ that depend on a continuous 
variable $t$ and satisfy the system of commutation relations, 
\cite{saveliev1, saveliev2},    
\ba
& & [X^+ (t), X^- (t^{\prime})] = S (t, t^{\prime}) H(t^{\prime}) ~, ~~~~ 
[H(t) , H(t^{\prime})] = 0 ~, \nonumber\\
& & [H(t) , X^{\pm} (t^{\prime})] = \pm K(t, t^{\prime}) X^{\pm} (t^{\prime}) ~.  
\ea
$K(t, t^{\prime})$ is called the Cartan kernel of the 
algebra and it generalizes the Cartan matrix $K_{ij}$ of 
simple Lie algebras to the continuous case, whereas the typical choice for $S$ is
$S(t, t^{\prime}) = \delta(t-t^{\prime})$, which corresponds to $\delta_{ij}$ in the 
case of discrete roots. This algebraic structure can be  
alternatively defined by smearing the Cartan-Weyl generators with arbitrary 
functions $\varphi (t)$ of compact support, as in the theory of distributions. 
Thus, setting  
\be
A(\varphi) = \int dt \varphi (t) A(t) 
\ee
the basic commutation relations assume the form 
\ba
& & [X^+ (\varphi), X^- (\psi)] = H(S(\varphi, \psi)) ~, ~~~~ 
[H(\varphi) , H(\psi)] = 0 ~, \nonumber\\
& & [H(\varphi) , X^{\pm} (\psi)] = \pm X^{\pm} (K(\varphi, \psi)) \label{bosalg}    
\ea

In general, the operators $K$ and $S$ are thought to be bilinear maps  
on the vector space of functions $\varphi (t)$ satisfying the relations   
\be
K(\varphi, K(\psi, \chi)) = K(\psi, K(\varphi, \chi)) ~, ~~~~~ 
S(\varphi, K(\psi, \chi)) = S(K(\psi, \varphi), \chi) 
\ee
for consistency with the Jacobi 
identities. Here, we only need to consider the case $K(\varphi , \psi) = (K\varphi) \psi$ 
and $S(\varphi, \psi) = S (\varphi \psi)$ prescribed by linear operators  
$K$ and $S$, which always ensure the validity of the Jacobi identities. The corresponding
continual Lie algebra will be denoted by ${\cal G}(K, S)$, but it can be easily shown that
${\cal G}(K, S) \simeq {\cal G}(\tilde{K}, 1)$ with $\tilde{K} = KS$; for it suffices
to define a new basis of Cartan generators $\tilde{H}(\varphi) = H(S \varphi)$. Therefore, 
for all practical purposes, the continual Lie algebras can be brought into standard  
form, which is abbreviated to ${\cal G}(K)$, 
setting $S=1$ and dropping the tilde from $K$. From now on we will use this notation 
unless stated otherwise. The Cartan kernel     
$K$ can be symmetrizable, but for the purposes of the present work we will also 
allow for algebras with anti-symmetric Cartan kernels $K(t, t^{\prime}) = 
- K(t^{\prime}, t)$. In fact, the geometric flows turn out to be naturally related to 
continual Lie algebras with anti-symmetric kernels, as will be shown later 
in detail. 

The continual Lie algebras 
${\cal G}(K)$ are infinite dimensional $Z$-graded algebras with
direct decomposition into subspaces  
\be
{\cal G}(K) = \oplus_{n \in Z} {\cal G}_n ~. 
\ee
The Cartan-Weyl generators $H$ and $X^{\pm}$ span the components ${\cal G}_0$ 
and ${\cal G}_{\pm 1}$, respectively, thus forming the local part of the algebra 
${\cal G}_{-1} \oplus {\cal G}_0 \oplus {\cal G}_{+1}$, whereas all other elements of it  
can be constructed recursively by taking successive commutators of the basic generators    
so that ${\cal G}_n = [{\cal G}_{n-1} , {\cal G}_{+1}]$ for all $n >1$ and 
${\cal G}_n = [{\cal G}_{n+1} , {\cal G}_{-1}]$ for all $n < -1$.  
The growth of the algebra which is characterized by the dimension of the 
subspaces ${\cal G}_n$ relative to the dimension of ${\cal G}_0$ can vary accordingly 
depending on the choice of $K$. The derivation of all commutation relations of the 
algebra ${\cal G}(K)$ is not an easy task as there is no known analogue of the 
Serre relations in the general case. As a result, the complete description of 
${\cal G}(K)$ is still lacking for arbitrary $K$,  
and in particular for those cases that will be encountered in the algebraic  
description of geometric flows. Nevertheless, the zero curvature formulation of the  
flows, and their integration, will be solely based on the local part of 
the algebras without ever requiring knowledge of the full structure of ${\cal G}(K)$ in 
either case.    

\subsection{Associated Toda systems}

The Toda system associated to any continual Lie algebra with Cartan operator $K$ is 
defined to be  
\be
\partial \bar{\partial} \Phi (z, \bar{z} ; t) = \int dt^{\prime} K(t^{\prime}, t) 
e^{\Phi (z , \bar{z} ; t^{\prime})} ~. \label{besme}  
\ee
$\Phi(z, \bar{z}; t)$ can be viewed as a one-parameter family of two dimensional fields 
that depend on a continuous variable $t$ rather than on discrete indices $i$, 
\cite{saveliev1, saveliev2}.   
As such, they generalize the Toda field equations associated to simple Lie algebras 
with Cartan matrix $K_{ij}$ to the case of infinite dimensional algebras with a 
continuous system of roots, at least formally. In analogy to ordinary Toda systems, this   
equation is cast into zero curvature form 
\be
[\partial + A_+ (z, \bar{z}), ~ \bar{\partial} + A_- (z, \bar{z}) ] = 0  
\ee
using appropriate gauge connections $A_{\pm}$ taking their values in the local part 
${\cal G}_0 \oplus {\cal G}_{\pm 1}$ of the algebra ${\cal G}(K)$. 

More precisely, let 
us consider the general ansatz for the two-dimensional gauge connections  
\be
A_{\pm} (z , \bar{z}) = H(u_{\pm}) + X^{\pm} (f_{\pm}) ~,  
\ee
which depend on functions $u_{\pm}$ and $f_{\pm}$ of $z$, $\bar{z}$ and $t$. 
Some of these functions are superfluous and they can be eliminated 
by employing the gauge invariance of the zero curvature condition under the 
transformations 
\be
A_+ \rightarrow g^{-1} (\partial + A_+) g ~, ~~~~~  
A_- \rightarrow g^{-1} (\bar{\partial} + A_-) g 
\ee
with arbitrary group elements $g = {\rm exp} H(\varphi)$. As it turns out, it is always 
possible to choose the gauge 
\be
A_+ = H(\Psi) + X^+(1) ~, ~~~~~ A_- = X^- (e^{\Phi}) 
\ee
and then use the commutation relations of the algebra in order to obtain the following  
system of equations from the zero curvature condition,   
\be
\bar{\partial} \Psi = e^{\Phi} ~, ~~~~~ \partial \Phi = K(\Psi) ~.  
\ee
Elimination of the function $\Psi$ leads to the continual Toda field equation 
\be
\partial \bar{\partial} \Phi = K(e^{\Phi}) ~, 
\ee
which is the smeared form of equation \eqn{besme} above.  

The general solution is obtained in analogy to ordinary Toda systems (for a review, 
see, for instance, \cite{saveliev3}), by 
introducing a normalized highest weight state $|t>$ that depends on the continuous  
variable $t$,  
\be
X^+ (t^{\prime}) |t> = 0 ~, ~~~~ <t| X^- (t^{\prime}) = 0 ~, ~~~~ 
H(t^{\prime}) |t> = \delta (t-t^{\prime}) |t> 
\ee
with $<t|t>= 1$. Then, the one-parameter family of two 
dimensional free fields with the decomposition   
\be
\Phi_0 (z, \bar{z}; t) = f (z; t) + \bar{f} (\bar{z}; t) 
\ee
following from $\partial \bar{\partial} \Phi_0 = 0$, is used to express the Toda field 
configurations in the form, \cite{saveliev1, saveliev2},  
\be
\Phi (z, \bar{z}; t) = \Phi_0 (z, \bar{z}; t) -  
K\left( {\rm log} <t| 
M_+^{-1} (z) M_- (\bar{z})|t>\right) , \label{genex} 
\ee
where $M_{\pm}$ are given by the path-ordered exponentials of the Lie algebra 
elements in ${\cal G}_{\pm 1}$, respectively,   
\ba
M_+ (z) & = & {\cal P} {\rm exp} \left(\int^{z} dz^{\prime} 
\int dt^{\prime} e^{f (z^{\prime}; t^{\prime})} X^+ (t^{\prime}) 
\right) , \nonumber\\ 
M_- (\bar{z}) & = & {\cal P} {\rm exp} \left(\int^{\bar{z}} d\bar{z}^{\prime} 
\int dt^{\prime} e^{\bar{f} (\bar{z}^{\prime}; t^{\prime})} X^- (t^{\prime}) 
\right) . 
\label{poex}  
\ea
This formula summarizes the B\"acklund transformations of continual Toda systems,  
thus providing a free field realization of their general solution.   

The details can be worked out by expanding $M_{\pm}$ in power series and compute the 
expectation value of their product in the highest weight state, as 
\ba
& & <t|M_+^{-1} M_- |t> = 1 + \sum_{n=1}^{\infty} (-1)^n \int^{z} dz_1 \cdots 
\int^{z_{n-1}} dz_n \int^{\bar{z}} d\bar{z}_1 \cdots \int^{\bar{z}_{n-1}} d\bar{z}_n 
\times \nonumber\\  
& & ~~~~~~ \times \int \prod_{i=1}^{n} dt_i \int \prod_{i=1}^{n} dt_i^{\prime} 
{\rm exp} f (z_i ; t_i) {\rm exp} \bar{f} (\bar{z}_i ; t_i^{\prime}) 
D_t^{\{t_1, \cdots , t_n ; t_1^{\prime} , \cdots , t_n^{\prime}\}} \label{foposee}  
\ea
with the ordering $z \geq z_1 \geq \cdots \geq z_{n-1} \geq z_n$ and likewise for the  
$\bar{z}$'s. 
The structure of the algebra ${\cal G}(K)$ is fully encoded into the elements 
\be
D_t^{\{t_1, t_2, \cdots , t_n ; t_1^{\prime} , t_2^{\prime} , \cdots , t_n^{\prime}\}} 
= <t|X^+(t_1) X^+(t_2) \cdots X^+(t_n) X^-(t_n^{\prime}) \cdots X^-(t_2^{\prime}) 
X^-(t_1^{\prime}) |t> 
\ee
that depend on the choice of Cartan operator $K$. They are determined using the recursive  
relations  
\be
D_t^{\{t_1, t_2, \cdots , t_n ; t_1^{\prime} , t_2^{\prime} , \cdots , t_n^{\prime}\}} 
= \sum_{j=1}^n \delta (t_n - t_j^{\prime}) \left(\delta (t-t_j^{\prime}) - 
\sum_{k=1}^{j-1} K(t_j^{\prime}, t_k^{\prime}) \right) 
D_t^{\{t_1, \cdots , t_{n-1} ; t_1^{\prime} , \cdots , \hat{t}_j^{\prime} , \cdots , 
t_n^{\prime}\}} 
\ee
with $D_t^{\{t_1; t_1^{\prime}\}} = \delta (t-t_1) \delta (t-t_1^{\prime})$. Here, 
$\hat{t}_j^{\prime}$ is used to denote that $t_j^{\prime}$ has been omitted by 
contraction with $t_m$ while  
shifting $X^{+}$'s to the right and $X^{-}$'s to the left.  

Thus, straightforward extension of the group theoretical methods used for the integration
of Toda systems for simple Lie algebras allows to construct the general solution of 
continual Toda field equations and obtain a formal infinite power series expansion around 
free field configurations. In regions where the exponential self-interaction becomes 
negligible one simply has $\Phi \simeq \Phi_0$ for the corresponding Toda fields. Such 
formal expansion may or may not converge in the strict mathematical sense, but there is 
no general way to decide on such issues at the moment.

The simplest and mostly studied example of this kind is provided by the so 
called heavenly equation, which 
is the following non-linear equation for a field $\Phi(z, \bar{z}; t)$, 
\be
{\partial^2 \over \partial t^2} e^{\Phi} = - \partial \bar{\partial} \Phi ~. 
\label{heavenl}
\ee
This equation is not parabolic, unlike the equations associated to geometric 
evolutions, for it involves second derivatives in space and time. It  
arises in the context of four-dimensional 
hyper-K\"ahler manifolds that admit a non-triholomorphic isometry associated to a 
rotational $S^1$ action, and describes the self-duality condition on the 
Riemann curvature tensor of the metric in adapted coordinates, 
\cite{das}. In the present context, 
it is identified with the Toda field equation for the 
large $N$ limit of the $SU(N)$ algebra, where the Cartan matrix   
$K_{ij} = 2 \delta_{i,j} - \delta_{i+1, j} - \delta_{i, j+1}$ is replaced by the 
kernel $K(t, t^{\prime}) = -\delta^{\prime \prime} (t-t^{\prime})$ using a 
continuous parameter $t$ to label the vertices of the Dynkin diagram, 
\cite{park}. Thus, the 
corresponding Cartan operator is $K = - \partial^2 / \partial t^2$.   
Integration of the heavenly equation by group theoretical methods has been carried out,  
as for Toda theories, and various geometrically interesting solutions have been 
studied explicitly; the examples include the Eguchi-Hanson, Taub-NUT and Atiyah-Hitchin 
metrics, which all admit at least one non-triholomorphic $S^1$ action. 
In that context, the validity of the formal power series expansion 
of the general solution in terms of free fields has been tested in several cases,  
and the interested reader is referred to the literature, \cite{sfe}, for details.

\subsection{Generalization to supercontinual Lie algebras}

Next, we introduce the more general class of supercontinual Lie algebras 
which are defined by the basic system of commutation relations
\ba
& & [X^+ ({\cal F}), X^- ({\cal G})] = H({\cal F} {\cal G}) ~, ~~~~ 
[H({\cal F}) , H({\cal G})] = 0 ~, \nonumber\\
& & [H({\cal F}) , X^{\pm} ({\cal G})] = \pm X^{\pm} ((K{\cal F}) {\cal G})   
\label{mostgene} 
\ea
for the generators $H$ and $X^{\pm}$. The difference with the ordinary case is that 
the smearing functions ${\cal F}$ are taken to be 
super-functions on $R^{1|1}$ superspace with 
coordinate $T=(t, \theta)$ having $\theta^2 = 0$. 
Likewise, the generators $H(T)$ and $X^{\pm}(T)$ are 
not bosonic but they also admit an expansion in $R^{1|1}$ superspace. However, 
the smeared form of the generators $H({\cal F})$ and $X^{\pm}({\cal F})$ is 
considered to be bosonic so that they possess commutation relations rather than 
anti-commutation relations. Thus, in effect, supercontinual Lie algebra provide 
a special class of infinite dimensional Lie algebras that can be brought into 
a standard Cartan form with the aid of a Grassmann variable $\theta$.    

The corresponding Cartan operator $K$ can be an ordinary linear operator which is  
defined on the space of functions of a continuous variable $t$ and acts separately 
on each component of the smearing 
super-functions ${\cal F}$. More generally, it can be a super-operator acting 
linearly on the space of super-functions ${\cal F}(T)$.  
In either case, the Cartan operator has to satisfy the relations  
\be
(K{\cal F}) {\cal G} = {\cal G}(K{\cal F}) ~, ~~~~~ (K{\cal F}) (K{\cal G}) = 
(K{\cal G}) (K{\cal F}) 
\ee
for all ${\cal F}$ and ${\cal G}$, 
as follow from the Jacobi identity for the two triplets $(H, X^+, X^-)$ and 
$(H, H, X^{\pm})$. These put no special restrictions on the choices 
that are available, and, therefore, the class of super-continual Lie algebras  
is well defined for all $K$.   
A particularly interesting example, which is relevant for the algebraic description 
of the Calabi flow, corresponds to the odd choice 
$K = \partial /\partial \theta - \theta \partial / \partial t$, as will be seen later.  
Other choices are also 
possible, but their physical relevance is not known in general.      

The smearing super-functions ${\cal F}$ can be expanded in components as 
\be
{\cal F}(T) = \varphi_0 (t) + \theta \varphi_1 (t) ~, \label{misufun}  
\ee
using the superspace coordinates $T= (t, \theta)$. Likewise, the generators 
of the algebra can be 
viewed as being parametrized by a continuous super-time 
parameter $T$ with respect to which 
they also admit expansion into components, 
\be
A(T) = A_0(t) + \theta A_1(t) ~, \label{misufun2}  
\ee
with $A$ being either $H$ or $X^{\pm}$. By definition, the two components  
$\varphi_0(t)$ and $\varphi_1(t)$ are even functions taking   
values in ordinary numbers, and 
similarly $A_0(t)$ and $A_1(t)$ are also even. The smearing is defined 
by complete integration in $R^{1|1}$ 
superspace with measure $dT = dt d\theta$, as follows,  
\be
A({\cal F}) = \int A(T) {\cal F}(T) dT = A_0(\varphi_1) + A_1(\varphi_0) ~,
\ee
using the known identities $\int d \theta = 0$ and $\int \theta d\theta = 1$. Thus, 
all such $A({\cal F})$ are even, as required.  
Then, the structure of the 
commutation relations can be worked out in components for the bosonic generators
$H_i(\varphi)$ and $X_i^{\pm}(\varphi)$ with $i=0,1$, depending on the form of $K$. 

Toda systems associated to such algebraic structures can be defined as usual 
by the zero curvature form $[\partial + A_+(z, \bar{z}), ~ 
\bar{\partial} + A_-(z, \bar{z})]=0$, choosing 
\be
A_+ = H({\cal G}) + X^+(1) ~, ~~~~~ A_- = X^-(e^{\cal F}) 
\ee
for the gauge connections. Then, if the super-functions ${\cal F}$ and ${\cal G}$ 
are related to each other as $\partial {\cal F} = K({\cal G})$, the 
corresponding Toda field equation will be 
\be
\partial \bar{\partial} {\cal F} = K(e^{\cal F}) ~.  
\ee
Expanding ${\cal F}$ in components and comparing the even and odd parts of the equation 
yields an equivalent description of the associated Toda system. The general 
solution can also be obtained in this case by straightforward generalization of 
the group theoretical methods developed for ordinary Toda systems. Further details 
will be given later for the particular example of the Calabi flow.    

Other possibilities may arise, in principle,  by considering ${\cal F}$ as  
super-functions on $R^{1|N}$ superspace, according to our general 
discussion. The structure of the associated supercontinual Lie algebras  
can also be worked out in component form, but become
more complicated for higher vales of $N$.

\section{Ricci flow}
\setcounter{equation}{0}

According to the general discussion in section 2,  
the Ricci flow equation in two dimensions assumes the following simple form in a   
system of conformally flat coordinates $(z, \bar{z})$,  
\be
\partial \bar{\partial}  
\Phi (z, \bar{z}; t) =   
\partial_t e^{\Phi (z, \bar{z}; t)} ~, \label{rfe}  
\ee
where $\Phi$ is the conformal factor of the metric undergoing continuous deformations. 
As such, it admits a natural algebraic interpretation in the framework of Toda 
field equations for the choice of continual Lie algebra with Cartan operator
$K = \partial / \partial t$. In particular,  
following \cite{bakas1, bakas2}, we consider the continual Lie algebra 
whose local part satisfies the commutation relations, in smeared form,   
\ba
& & [X^+ (\varphi), X^- (\psi)] = H(\varphi \psi) ~, ~~~~ 
[H(\varphi) , H(\psi)] = 0 ~, \nonumber\\
& & [H(\varphi) , X^{\pm} (\psi)] = \pm X^{\pm} (\varphi^{\prime} \psi) ~.     
\ea
Then, the two-dimensional Ricci flow, which is viewed as Toda system for  
this algebra, \cite{bakas1, bakas2}, admits the zero curvature formulation
\be
[\partial + A_+ (z, \bar{z}), ~ \bar{\partial} + A_- (z, \bar{z}) ] = 0 ~, 
\ee
where $A_{\pm}$ take values in the local part as 
\be
A_+ = H(\Psi) + X^+ (1) ~, ~~~~~ A_- = X^- (e^{\Phi}) ~.  
\ee
Also, following the general discussion in section 3, we find that  
$\partial \Phi = \partial_t \Psi$ and $\bar{\partial} \Psi = {\rm exp} \Phi$ 
and, hence, $\Phi$ satisfies the Ricci flow equation 
\eqn{rfe} after eliminating $\Psi$.   

The main ingredient in this particular formulation of the Ricci flow 
is the use of a continual Lie algebra to incorporate  
the deformation variable $t$ into its root system. Thus, the equation is 
integrable in two dimensions using flat space coordinates $(z, \bar{z})$, 
without reaching a contradiction of terms with the  
dissipative behavior in time. 
In this context, the parabolic nature of the flow,    
which is first order in $t$, is held responsible for the 
novel structure of the underlying 
Lie algebra with anti-symmetric 
Cartan kernel
\be
K(t, t^{\prime}) = - {\partial \over \partial t} \delta (t-t^{\prime}) ~. 
\ee
Such algebras have been studied very little in the mathematics literature, but it has already
been noted that ${\cal G}(\partial / \partial t)$ 
exhibits exponentially fast growth beyond its local part. As a 
result, its complete structure is still lacking and there is no framework in which to 
realize transformation rules for all the generators. Also, the precise meaning of integrability
for systems associated to such algebras should be understood better in the future, and in 
connection with the explicit construction of their conservation laws in 
two-dimensional space. 
The fact is that they
become relevant for the algebraic description of certain dynamical problems with the Ricci
flow being the first example of this kind. The Calabi flow also falls in this category  
for appropriate choice of Cartan operator in the class of supercontinual Lie 
algebras, as will be seen later in detail.  

Next, we apply the general formalism to describe the solutions of the two-dimensional Ricci
flow equation in terms of a one-parameter family of free fields 
$\Phi_0 (z, \bar{z}; t) = f (z; t) + \bar{f} (\bar{z}; t)$. The general expression \eqn{genex} 
specializes to  
\be
\Phi (z, \bar{z}; t) = \Phi_0 (z, \bar{z}; t) - \partial_t \left({\rm log} <t| 
M_+^{-1} (z) M_- (\bar{z})|t> \right) ,  
\ee
where $M_{\pm}$ are the path ordered exponentials \eqn{poex} that should be 
evaluated for the particular algebra 
${\cal G}(\partial / \partial t)$. This provides the solution in  
closed form by group theoretical methods, but it is very implicit. In practice, one should
evaluate recursively the elements 
$D_t^{\{t_1, t_2, \cdots , t_n ; t_1^{\prime} , t_2^{\prime} , \cdots , t_n^{\prime}\}}$  
and substitute them into the general power series expansion \eqn{foposee}.  

Explicit calculation yields in this case the following results, up to $n=3$, \cite{bakas1},  
\ba
& & D_t^{\{t_1; t_1^{\prime}\}} = \delta(t-t_1) \delta (t_1 - t_1^{\prime}) ~, 
\nonumber\\
& & D_t^{\{t_1, t_2; t_1^{\prime}, t_2^{\prime}\}} = \delta(t-t_1) 
\delta (t_1 - t_1^{\prime}) \delta (t_2 - t_2^{\prime})\left(2 \delta(t-t_2) 
- \partial_t \delta(t-t_2) \right) , \nonumber\\
& & D_t^{\{t_1, t_2, t_3; t_1^{\prime}, t_2^{\prime}, t_3^{\prime}\}} = \delta(t-t_1) 
\delta (t_1 - t_1^{\prime}) \delta (t-t_3) \delta (t_2 - t_3^{\prime}) 
\delta (t_3 - t_2^{\prime}) \times \nonumber\\
& & ~~~~~~~~~~ \times \left(2\delta (t-t_2) - \partial_t \delta (t-t_2) \right) + 
\nonumber\\
& & ~~~~~~ + \delta (t-t_1) \delta (t_1 - t_1^{\prime}) \delta (t_2 - t_3^{\prime}) 
\delta (t_3 - t_2^{\prime}) \times \nonumber\\
& & ~~~~~~~~~~ \times \left(\delta (t-t_3) - \partial_t \delta (t-t_3) \right) 
\left(2\delta (t-t_2) - \partial_t \delta (t-t_2) \right) + \nonumber\\
& & ~~~~~~ + \delta (t-t_1) \delta (t_1 - t_1^{\prime}) \delta (t_2 - t_2^{\prime}) 
\delta (t_3 - t_3^{\prime}) \times \nonumber\\
& & ~~~~~~~~~~ \times \left(2\delta(t-t_2) - \partial_t \delta (t-t_2) \right) 
\left(\delta(t-t_3) - \partial_t \delta (t-t_3) - \partial_{t_2} \delta (t_2 - t_3)
\right)  \label{expforcomp}  
\ea
and so on for higher $n$. The expressions become considerably longer beyond $n=3$ 
and they are not included here.  

In special cases, where the corresponding 
geometric deformations are axially symmetric, the solutions can be taken to depend only on the 
combination of coordinates $z + \bar{z} := Y$. Then, for free field configurations of the 
form $\Phi_0 (Y; t) = cY + d(t)$, the systematic expansion of the path ordered exponential
leads to the power series solution  
\ba
\Phi & = & \Phi_0 + {1 \over (1! ~c)^2} \partial_t e^{\Phi_0} + {1 \over (2! ~c^2)^2} 
\partial_t \left(e^{\Phi_0} \partial_t e^{\Phi_0} \right) + {1 \over (3! ~c^3)^2} 
\partial_t \left( 3e^{\Phi_0} (\partial_t e^{\Phi_0})^2 + e^{2\Phi_0} \partial_t^2 
e^{\Phi_0} \right) \nonumber\\
& & + {1 \over (4! ~c^4)^2} \partial_t \left(e^{3\Phi_0} \partial_t^3 e^{\Phi_0} + 
17 e^{2\Phi_0} (\partial_t e^{\Phi_0})(\partial_t^2 e^{\Phi_0}) + 18 e^{\Phi_0} 
(\partial_t e^{\Phi_0})^3 \right) \nonumber\\
& & + {1 \over (5! ~c^5)^2} \partial_t \left(e^{4\Phi_0} \partial_t^4 e^{\Phi_0} + 
36 e^{3\Phi_0} (\partial_t e^{\Phi_0})(\partial_t^3 e^{\Phi_0}) + 35 e^{3\Phi_0} 
(\partial_t^2 e^{\Phi_0})^2 \right. \nonumber\\
& & \left. ~~~~~~~~~~~~~~~~~ + 324 e^{2\Phi_0} (\partial_t e^{\Phi_0})^2 
(\partial_t^2 e^{\Phi_0}) + 180 e^{\Phi_0} (\partial_t e^{\Phi_0})^4 \right)  
+ \cdots  \label{fitexp}  
\ea
by evaluating the contribution of all terms in the expansion up to order $n=5$ included.
It should be understood as a power series expansion of ${\rm exp} \Phi$ around the free field 
configuration ${\rm exp}\Phi_0$ that becomes very small when $Y \rightarrow \pm \infty$
for $c<0$ or $>0$, respectively.  
It can also be verified directly that this expansion satisfies the Ricci flow equation
order by order in powers of ${\rm exp} \Phi_0$. 

The validity of the formal power series expansion and its convergence have be tested in  
various examples of axially symmetric geometric deformation that are known in closed form
using more direct (mini-superspace) methods. A notable example is the sausage model that
corresponds to the choice of conformal factor, \cite{sausa},  
\be
e^{\Phi(Y; t)} = {2 \over a(t) + b(t) {\rm cosh}2Y} ~~~~ {\rm with} ~~~~ 0 \leq X \leq 2\pi ~,  
~~~~ -\infty < Y < +\infty ~,  
\ee
while $t$ is running from $-\infty$ to $0$ and  
\be
a(t) = - \gamma {\rm coth}(2\gamma t) ~, ~~~~~ b(t) = -{\gamma \over {\rm sinh} (2\gamma t)} ~.  
\ee
Note that $a \geq b \geq 0$ since $t$ assumes only negative values; $t \rightarrow -\infty$ 
corresponds to the asymptotically free ultra-violet limit of the $O(3)$ sigma model in 
two dimensions. This solution  
describes a family of axially symmetric deformations of the round 2-sphere for all values
of the parameter $\gamma \geq 0$. It fits the expansion \eqn{fitexp} in power of 
${\rm exp}(-2Y)$ for large $Y$, \cite{bakas1}, provided that the 
free field is chosen to be $\Phi_0(Y;t) = cY + d(t)$ with
\be
c=-2 ~, ~~~~~ d(t) = {\rm log} \left(-{4 \over \gamma} {\rm sinh}(2\gamma t) \right) .  
\ee
   
Thus, there is mounting confidence that the formal manipulations used for the integration of the 
Ricci flow indeed make good sense. 
Further details can be found in the published works, \cite{bakas1, bakas2}. 
The only drawback of the present formalism is the inability 
to prescribe directly the geometric shapes of space from different choices of free
fields.     

\section{Calabi flow}
\setcounter{equation}{0}

Next, we turn to the two-dimensional Calabi flow, which will be shown to admit zero 
curvature formulation for appropriately chosen gauge connections. The method of 
investigation involves the introduction of anti-commuting variables following 
a simple observation that formally relates it to the Ricci flow. Then, we give 
full justification for the choice of the 
infinite dimensional algebra that enables to cast 
the Calabi flow in integrable form, as Toda system, and extend the validity of 
the group theoretical methods introduced earlier in order to describe (at least 
formally) the general solution in terms of free fields, as for the Ricci flow.    
The results described below complement earlier work on the subject by providing 
full justification of the seemingly ad hoc choices made in reference \cite{bakala}
and go far beyond it in some directions. 

\subsection{Calabi flow as super-evolution}
 
According to the general discussion found  
in section 2, the two-dimensional Calabi flow assumes explicitly the form 
\be
\partial_t e^{\Phi} = - \partial \bar{\partial} \left(e^{-\Phi} 
\partial \bar{\partial} \Phi \right) \label{cfe}  
\ee
in a system of conformally flat coordinates with conformal factor 
$\Phi(z, \bar{z}; t)$ varying with time $t$. Specializing an earlier observation,   
it is instructive to compared it to the two-dimensional 
Ricci flow equation \eqn{rfe}, which is 
equivalently stated as 
\be
{\partial^2 \over \partial t^2} e^{\Phi} = \partial \bar{\partial} \left(e^{-\Phi} 
\partial \bar{\partial} \Phi \right)  
\ee
after taking its derivative with respect to $t$. Indeed, the Calabi flow appears to  
be formally related to the Ricci flow by taking the square root of its evolution  
operator $\partial / \partial t$, up to a sign.  

This simple correspondence motivates us to introduce an anti-commuting variable $\theta$,
with $\theta^2 = 0$, as  
the supersymmetric partner of the deformation variable $t$ and write down the following 
abstract equation  
\be
\partial \bar{\partial} {\cal F} = {\cal D}_T {\rm exp} {\cal F} \label{suevol}  
\ee
where 
\be
{\cal F} (z, \bar{z}; t, \theta) = \Phi (z, \bar{z}; t) + 
\theta \Psi (z, \bar{z} ; t) 
\label{cfe2}
\ee
is a {\em mixed} superfield with bosonic components $\Phi$, $\Psi$ and 
\be
{\cal D}_T = {\partial \over \partial \theta} - \theta {\partial \over \partial t} 
\ee
is the associated super-derivative in $R^{1|1}$ superspace with coordinates $T=(t, \theta)$.
It satisfies the relation  
${\cal D}_T^2 = - \partial / \partial t$, whereas the complementary 
super-derivative operator $\partial/ \partial \theta + \theta \partial/  
\partial t$ squares to $\partial/ \partial t$, since it relates $t$ to $-t$.   
The equation above can be viewed as some short of super-evolution equation of 
Ricci type with respect to the super-time $T$.
Its precise content can be extracted by working in components. 
Expanding ${\rm exp}{\cal F}  = (1 + \theta \Psi) {\rm exp} \Phi$ and comparing the
even and odd parts, it readily follows that   
\be
e^{\Phi} \Psi = \partial \bar{\partial} \Phi ~, ~~~~~ 
\partial_t e^{\Phi} = -\partial \bar{\partial} \Psi ~. \label{altsyst}  
\ee
Subsequently, by eliminating $\Psi$, 
it follows that $\Phi$ satisfies the Calabi flow equation
\eqn{cfe} above. Thus, by taking the square root of the time derivative operator
allows to connect the two distinct classes of geometric deformations of 
second and fourth, respectively, via super-evolution, as was anticipated before 
on general grounds.

This simple result has far reaching consequences, as it allows to prove integrability  
of the two-dimensional Calabi flow by viewing it as Toda field equation for
a particular supercontinual algebra with odd Cartan operator $K={\cal D}_T$, 
in analogy to the Ricci flow that corresponds to the choice $K=\partial/ \partial t$.
More precisely, we are led to introduce an algebra  
whose local part satisfies the commutation relations   
\ba
& & [X^+ ({\cal F}), X^- ({\cal G})] = H({\cal F} {\cal G}) ~, ~~~~ 
[H({\cal F}) , H({\cal G})] = 0 ~, \nonumber\\
& & [H({\cal F}) , X^{\pm} ({\cal G})] = 
\pm X^{\pm} (({\cal D}_T {\cal F}) {\cal G}) ~. \label{basycr}     
\ea
It defines a supercontinual Lie algebra in the nomenclature introduced earlier.  

Then, following the general discussion in section 3, we write down the zero
curvature condition  
\be
[\partial + A_+ (z, \bar{z}), ~ \bar{\partial} + A_- (z, \bar{z}) ] = 0 ~, 
\ee
where the connections $A_{\pm}$ take values in the local part of the algebra, as  
\be
A_+ = H({\cal G}) + X^+ (1) ~, ~~~~~ A_- = X^- ({\rm exp}{{\cal F}}) ~. \label{gaucon}   
\ee
The superfield ${\cal F}$ satisfies the flow  
equation \eqn{suevol}, provided that ${\cal G}$ is constrained to 
\be
\partial {\cal F} = 
{\cal D}_T {\cal G} \label{congau}  
\ee
by the zero curvature condition. Thus, the Calabi flow is indeed 
an integrable two-dimensional system of Toda type. 

Next, the structure of the algebra \eqn{basycr} is analyzed in more detail, 
using components in $R^{1|1}$ superspace, and then used further to 
integrate the equation by group theoretical methods, as in ordinary Toda systems.   

\subsection{Working in components}

The algebra introduced for the Calabi flow 
is not a supersymmetric extension of what was used before  
for the Ricci flow, but it is bosonic. 
Our prescription did not involve any substitution of the 
space derivatives $\partial$ and $\bar{\partial}$ by the supersymmetric 
covariant derivatives ${\cal D}_{\pm}$, respectively,
as it is customary in supersymmetric systems, but only used the odd generalization of the
time evolution operator in $R^{1|1}$ superspace, in order to converge a second order 
differential equation in $(z, \bar{z})$ into fourth order. 
It can be alternatively viewed as changing the 
kernels of the algebra for the Ricci flow,  
\be
S(t, t^{\prime}) = \delta (t-t^{\prime}) ~, ~~~~~ 
K(t, t^{\prime}) = - \partial_t \delta (t-t^{\prime}) ~,   
\ee
into the two super-kernels, respectively,  
\ba
&  & S(T, T^{\prime}) = \delta (T - T^{\prime}) = (\theta - \theta^{\prime}) 
\delta (t-t^{\prime}) ~, \\  
&  & K(T, T^{\prime}) = - {\cal D}_T \delta (T-T^{\prime}) = - \delta (t-t^{\prime}) 
- \theta \theta^{\prime} \partial_t \delta (t-t^{\prime}) ~,  
\ea
thus giving rise to the bosonic supercontinual Lie algebra with $K={\cal D}_T$.  
Of course, the passage from smeared to unsmeared form of the algebra has to
be done with care, since the integration over the Grassmann coordinates 
$\theta$ and $\theta^{\prime}$ will flip signs in some places. This generalization    
also changes the symmetry of the corresponding kernels, since $S(T, T^{\prime})$  
is now anti-symmetric and $K(T, T^{\prime})$ is symmetric under the interchange 
of their arguments, $T \leftrightarrow T^{\prime}$. 

It is useful in many respects to work out the form of the algebra in components  
by expanding the generators,  
\be
X^{\pm}(T) = X_0^{\pm}(t) + \theta X_1^{\pm}(t) ~, ~~~~~ 
H(T) = H_0(t) + \theta H_1(t) ~.   
\ee
All components $X_i^{\pm}(t)$ and $H_i(t)$ with $i=0,1$ are bosonic, and one  
arrives to the equivalent description of the local part of the algebra with commutation 
relations  
\ba
& & [H_0(t), ~ X_0^{\pm}(t^{\prime})] = \pm \delta (t-t^{\prime}) X_0^{\pm} (t^{\prime}) ~, 
~~~~ [X_1^{\pm}(t), ~ X_0^{\mp}(t^{\prime})] = \pm \delta (t-t^{\prime}) H_0 (t^{\prime}) ~, 
\nonumber\\
& & [H_0(t), ~ X_1^{\pm}(t^{\prime})] = \pm \delta (t-t^{\prime}) X_1^{\pm} (t^{\prime}) ~, 
~~~~ [X_1^+ (t), ~ X_1^- (t^{\prime})] = \delta (t-t^{\prime}) H_1 (t^{\prime}) ~, 
\nonumber\\
& & [H_1(t), ~ X_1^{\pm}(t^{\prime})] = \pm \partial_t \delta (t-t^{\prime}) 
X_0^{\pm} (t^{\prime}) ~, 
\ea
whereas the remaining ones are trivial, i.e., 
\ba
& & [H_1(t), ~ X_0^{\pm} (t^{\prime})] = 0 ~, ~~~~ [X_0^+ (t), ~ X_0^{-} (t^{\prime})] 
= 0 ~, \nonumber\\
& & [H_i (t), ~ H_j (t^{\prime})] = 0 ~. 
\ea

Smearing with ordinary functions of the continuous variable $t$, the algebra takes the
form  
\ba
& & [H_0(\varphi), ~ X_0^{\pm}(\psi)] = \pm X_0^{\pm} (\varphi \psi) ~, 
~~~~ [X_1^{\pm}(\varphi), ~ X_0^{\mp}(\psi)] = \pm H_0 (\varphi \psi) ~, 
\nonumber\\
& & [H_0(\varphi), ~ X_1^{\pm}(\psi)] = \pm X_1^{\pm} (\varphi \psi) ~, 
~~~~ [X_1^+ (\varphi), ~ X_1^- (\psi)] = H_1 (\varphi \psi) ~, 
\nonumber\\
& & [H_1(\varphi), ~ X_1^{\pm}(\psi)] = 
\mp X_0^{\pm} (\varphi^{\prime} \psi) ~, 
\ea
and 
\ba
& & [H_1(\varphi), ~ X_0^{\pm} (\psi)] = 0 ~, ~~~~ [X_0^+ (\varphi), ~ X_0^{-} (\psi)] 
= 0 ~, \nonumber\\
& & [H_i (\varphi), ~ H_j (\psi)] = 0 ~. 
\ea
One could have started directly from such an algebra, without prior justification, 
as in reference \cite{bakala}, and 
use it to establish the integrability of the Calabi flow\footnote{There is a slight 
difference of signs in the commutation relations of the algebra and those appearing 
in reference \cite{bakala}. The two choices are simply related to each other by changing
$X_0^{\pm} \rightarrow i X_0^{\pm}$ and $X_1^{\pm} \rightarrow \mp i X_1^{\pm}$ and 
they do not affect the proof of integrability. The 
signs that arise here guarantee that the algebra is compatible with the Hermiticity 
condition $H_i^{\dagger} = H_i$ and ${X_i^{\pm}}^{\dagger} = X_i^{\mp}$ for all 
$i=0, 1$, unlike the other case. These Hermiticity conditions turn out to 
be essential for the actual integration of the equation by group theoretical 
methods using highest weight representations, as will be seen later.}. 
However, employing the notion
of supercontinual Lie algebras proves useful for putting its structure in neat form 
and making direct connection with the special class of integrable systems of Toda type.  
 
The zero curvature formulation of the Calabi flow based on the algebra \eqn{basycr} can  
also be analyzed in components. The gauge connections \eqn{gaucon} take the form  
\be
A_+ = H_0 (f) + H_1(g) + X_1^+(1) ~, ~~~~ A_- = X_0^-(\Psi e^{\Phi}) + X_1^-(e^{\Phi}) ~, 
\ee
using the expansion of the mixed superfields ${\cal F} = \Phi + \theta \Psi$ and  
${\cal G} = g + \theta f$.  
Also, the relation \eqn{congau} among these fields reads in components as
\be
\partial \Phi = f ~, ~~~~~ \partial \Psi = -\partial_t g ~. 
\ee
As a result, the zero curvature condition implies two more equations that follow 
by comparing 
coefficients of the terms $H_0$ and $H_1$, respectively,  
\be
\bar{\partial} f = \Psi e^{\Phi} ~, ~~~~~ \bar{\partial} g = e^{\Phi} ~,  
\ee
and lead to the system \eqn{altsyst} that accounts for the Calabi flow, after
eliminating all other functions but $\Phi$.

\subsection{Integration of Calabi flow}

The general solution can be obtained in analogy to the Ricci flow equation, as in ordinary 
Toda systems, by 
introducing a normalized highest weight state $|T>$, assuming that it exists, 
which depends on the continuous  
coordinate $T$ of $R^{1|1}$ superspace with $<T|T>=1$.  
Equivalently, in terms of components $(t, \theta)$, we may assume that 
the highest weight state is expanded as 
\be
|T> = |t>_0 + \theta |t>_1 
\ee
with the orthonormalization conditions that the bosonic state $|t>_0$ has unit 
norm and it is perpendicular to $|t>_1$. It is more convenient to present 
its defining properties in smeared form\footnote{Note that we systematically 
avoid to write the commutation relations of the supercontinual Lie algebra and 
the definition of its highest weight state starting from the unsmeared form of 
the super-generators $A(T)$ labeled by $T$ in $R^{1|1}$. In this way, various 
sign discrepancies associated to the ambiguity of ordering products of 
super-functions defined at different points of super-space are avoided, as 
they can easily lead to errors.},  
\be
X^+ ({\cal F}) |{\cal G}> = 0 ~, ~~~~ <{\cal G}| X^- ({\cal F}) = 0 ~, ~~~~ 
H({\cal F}) |{\cal G}> = |{\cal F} {\cal G}> ~, \label{primstate}  
\ee
where $A({\cal F})= A_0(\varphi_1) + A_1(\varphi_0)$ for the generators of 
the algebra and 
\be
|{\cal F}> = \int |T> {\cal F}(T) dT = \int \left(\varphi_1(t) |t>_0 + 
\varphi_0(t) |t>_1 \right) dt = |\varphi_1>_0 + |\varphi_0>_1
\ee
for the highest weight state, using the smearing super-function 
${\cal F}(T) = \varphi_0(t) + \theta \varphi_1(t)$.    

The components of equation \eqn{primstate} satisfy the smeared relations  
\ba
& & H_0 (\varphi) |\psi>_0 = 0 ~, ~~~~~~ H_0(\varphi) |\psi>_1 = 
|\varphi \psi>_0 ~, \nonumber\\
& & H_1 (\varphi) |\psi>_i = |\varphi \psi>_i ~~~~ {\rm for} ~~ 
i=0, 1 ~,  
\ea
and 
\be
X_0^+(\varphi) |\psi>_i = 0 = X_1^+(\varphi) |\psi>_i ~~~~ {\rm for} ~~ i=0,1 
\ee
together with their Hermitian conjugate equations,   
using ordinary functions $\varphi (t)$ and $\psi (t)$. The same relations can 
also be written in unsmeared bosonic form, as
\ba
& & H_0 (t^{\prime}) |t>_0 = 0 ~, ~~~~~~ H_0(t^{\prime}) |t>_1 = \delta (t-t^{\prime}) 
|t>_0 ~, \nonumber\\
& & H_1 (t^{\prime}) |t>_i = \delta (t-t^{\prime}) |t>_i ~~~~ {\rm for} ~~ 
i=0, 1 ~,  
\ea
and 
\be
X_0^+(t^{\prime}) |t>_i = 0 = X_1^+(t^{\prime}) |t>_i ~~~~ {\rm for} ~~ i=0,1 ~.  
\ee

It follows immediately from above that $H_0(t^{\prime}) |T> = 
\theta \delta (t-t^{\prime}) |t>_0$ and $H_1(t^{\prime}) |T> = \delta(t-t^{\prime}) 
|T>$, which will be useful later for the calculations. These relations also 
imply, in particular, the following condition for the unsmeared action of 
the Cartan generator on the highest weight state  
$H(T^{\prime}) |T> = (\theta + \theta^{\prime}) \delta (t-t^{\prime}) |T>$, which 
is neither $\delta(T-T^{\prime}) |T>$ nor $|T> \delta (T-T^{\prime})$ as  
compared to naive expectation.    

The zero curvature formulation of the Calabi flow, which is better stated as 
super-evolution equation \eqn{suevol}, allows us to express its 
general solution in superfield form 
\be
{\cal F} (z, \bar{z}; t, \theta) = {\cal F}_0 (z, \bar{z}; t) -    
{\cal D}_T \left( {\rm log} <T| 
M_+^{-1} (z) M_- (\bar{z})|T>\right) , \label{carim}  
\ee
where ${\cal F}_0 = \Phi_0 + \theta \Psi_0$ is a one-parameter family of 
mixed superfields whose components are two-dimensional bosonic free fields
that depend on the continuous variable $t$,   
\be
\Phi_0 (z, \bar{z}; t) = f (z; t) + \bar{f} (\bar{z}; t) ~, ~~~~~
\Psi_0 (z, \bar{z}; t) = \psi (z; t) + \bar{\psi} (\bar{z}; t) ~.  
\ee
Also, setting ${\cal F}_0(z; T) = f(z; t) + 
\theta \psi(z; t)$ and $\bar{\cal F}_0 (\bar{z}; t) = 
\bar{f} (\bar{z}; t) + \theta \bar{\psi} (\bar{z}; t)$, 
the operators $M_{\pm}$ are given by the following path-ordered exponentials  
\ba
M_+ (z) 
& = & {\cal P} {\rm exp} \left(\int^{z} dz^{\prime} 
\int e^{{\cal F}_0 (z^{\prime}; T^{\prime})} 
X^+ (T^{\prime}) dT^{\prime} \right) , \nonumber\\ 
M_- (\bar{z}) 
& = & {\cal P} {\rm exp} \left(\int^{\bar{z}} d\bar{z}^{\prime} 
\int e^{\bar{\cal F}_0 (\bar{z}^{\prime}; T^{\prime})} 
X^- (T^{\prime}) dT^{\prime} \right) .  
\ea

Performing the integration over the anti-commuting variable $\theta^{\prime}$ that 
enters in the exponent, one arrives at the equivalent expressions in terms of 
components, respectively,
\ba
M_+ (z) 
& = & {\cal P} {\rm exp} \left(\int^{z} dz^{\prime} 
\int dt^{\prime} e^{f (z^{\prime}; t^{\prime})} [X_1^+ (t^{\prime}) + 
\psi (z^{\prime} ; t^{\prime}) X_0^+ (t^{\prime})] \right) ,  
\nonumber\\
M_- (\bar{z}) 
& = & {\cal P} {\rm exp} \left(\int^{\bar{z}} d\bar{z}^{\prime} 
\int dt^{\prime} e^{\bar{f} (\bar{z}^{\prime}; t^{\prime})} 
[X_1^- (t^{\prime}) + \bar{\psi} (\bar{z}^{\prime}; t^{\prime}) 
X_0^- (t^{\prime})] \right) . \label{mgefpoes}  
\label{ano} 
\ea   
Then, formula \eqn{carim} summarizes the B\"acklund transformation of the Calabi flow, as 
a continual Toda system, and provides realization of its general solution
for the arbitrary free field configuration $(\Phi_0, \Psi_0)$. The solution for 
$\Phi (z, \bar{z}; t)$ is obtained by comparing the even terms of the mixed superfield 
equation \eqn{carim}, whereas $\Psi(z, \bar{z}; t)$ is obtained from the odd terms; 
the resulting expressions, which are only implicitly defined here, should satisfy the
equation $ \Psi = {\rm exp}(-\Phi) \partial \bar{\partial} \Phi$, by construction. 
The (formal) power series expansion of 
the solution follows by expanding the path-ordered exponentials, as usual. 

In the following we will 
consider first, for simplicity, the class of solutions with $\Psi_0 (z, \bar{z}; t) = 0$,
as they also resemble the solutions of the Ricci flow equation in some respect.  
The possibility to include free field configurations with $\Psi_0 \neq 0$ will be considered
separately, at the end of this section, as they lead to more complicated expressions for the 
free field expansion of the general solution. 

{\bf (i) Solutions with $\Psi_0 = 0$}: In this case, the path-ordered exponentials  
\eqn{ano} simplify considerably, as they involve only the generators $X_1^{\pm}(t^{\prime})$ 
in their exponent.   
Then, the detailed structure of the solution can be worked out in 
power series, which assumes the special form   
\ba
& & <T|M_+^{-1} M_- |T> = 1 + \sum_{n=1}^{\infty} (-1)^n \int^{z} dz_1 \cdots 
\int^{z_{n-1}} dz_n \int^{\bar{z}} d\bar{z}_1 \cdots \int^{\bar{z}_{n-1}} 
d\bar{z}_n 
\times \nonumber\\  
& & ~~~~~~ \times \int \prod_{i=1}^{n} dt_i \int \prod_{i=1}^{n} dt_i^{\prime} 
{\rm exp} f (z_i; t_i) {\rm exp} \bar{f} (\bar{z}_i; t_i^{\prime}) 
D_T^{\{t_1, \cdots , t_n ; t_1^{\prime} , \cdots , t_n^{\prime}\}} 
\ea
with the usual path-ordering prescription  
$z \geq z_1 \geq \cdots \geq z_{n-1} \geq z_n$, and likewise for the $\bar{z}$'s. 
The structure of the corresponding algebra \eqn{basycr} is fully encoded into the elements 
\be
D_T^{\{t_1, t_2, \cdots , t_m ; t_1^{\prime} , t_2^{\prime} , \cdots , t_m^{\prime}\}} 
= <T|X_1^+(t_1) X_1^+(t_2) \cdots X_1^+(t_n) X_1^-(t_n^{\prime}) \cdots X_1^-(t_2^{\prime}) 
X_1^-(t_1^{\prime}) |T> \label{matele}  
\ee
that only involve strings of the component operators $X_1^{\pm}(t)$, for $\Psi_0 =0$. 
Of course, even in this case, the other components of the algebra generators, 
$X_0^{\pm}(t)$ as well as  
$H_i(t)$, will also appear in the process of evaluating the corresponding 
matrix elements by pushing 
$X^+$'s to the right and $X^-$'s to the left, as they follow 
from the commutation relations. 
 
These elements can be computed recursively, as usual, starting from the defining relation 
$<T|T>=1$. 
The result turns out to be
\ba
& & D_T^{\{t_1; t_1^{\prime}\}} = \delta(t-t_1) \delta (t_1 - t_1^{\prime}) ~, 
\nonumber\\
& & D_T^{\{t_1, t_2; t_1^{\prime}, t_2^{\prime}\}} = \delta(t-t_1) 
\delta (t_1 - t_1^{\prime}) \delta (t_2 - t_2^{\prime})\left(2 \delta(t-t_2) 
+ \theta \partial_t \delta(t-t_2) \right) , \nonumber\\
& & D_T^{\{t_1, t_2, t_3; t_1^{\prime}, t_2^{\prime}, t_3^{\prime}\}} = \delta(t-t_1) 
\delta (t_1 - t_1^{\prime}) \delta (t-t_3) \delta (t_2 - t_3^{\prime}) 
\delta (t_3 - t_2^{\prime}) \times \nonumber\\
& & ~~~~~~~~~~ \times \left(2\delta (t-t_2) + \theta \partial_t \delta (t-t_2) \right) + 
\nonumber\\
& & ~~~~~~ + \delta (t-t_1) \delta (t_1 - t_1^{\prime}) \delta (t_2 - t_3^{\prime}) 
\delta (t_3 - t_2^{\prime}) \times \nonumber\\
& & ~~~~~~~~~~ \times \left(\delta (t-t_3) + \theta \partial_t \delta (t-t_3) \right) 
\left(2\delta (t-t_2) + \theta \partial_t \delta (t-t_2) \right) + \nonumber\\
& & ~~~~~~ + \delta (t-t_1) \delta (t_1 - t_1^{\prime}) \delta (t_2 - t_2^{\prime}) 
\delta (t_3 - t_3^{\prime}) \times \nonumber\\
& & ~~~~~~~~~~ \times \left(2\delta(t-t_2) + \theta \partial_t \delta (t-t_2) \right) 
\left(\delta(t-t_3) + \theta \partial_t \delta (t-t_3) + 
\theta \partial_{t_2} \delta (t_2 - t_3)
\right) - \nonumber\\
& & ~~~~~~ - \delta (t-t_1) \delta (t-t_2) \delta (t_1 - t_1^{\prime}) 
\delta (t_2 - t_2^{\prime}) \delta (t_3 - t_3^{\prime}) 
\partial_t \delta (t-t_3)   
\ea
and so on. Higher order terms with $n>3$ are also calculable, but their expressions 
turn out to be quite lengthy with rapidly increasing complexity. For this reason they are not 
appended here, although contributions up to $n=4$ and $5$ will be included in the 
subsequent evaluation of the field $\Phi$. 

The expressions above should also 
be compared to the matrix elements \eqn{expforcomp} corresponding to the Ricci flow, 
as they show  
many similarities for small values of $n$. Observe that the time derivatives of delta 
functions appearing in the Ricci flow matrix elements now come multiplied with $-\theta$, 
which is consistent with the change of $\partial / \partial t$ to 
${\cal D}_T = \partial / \partial \theta - \theta \partial / \partial t$ as one passes 
to the algebra of the Calabi flow. However, there is 
also a new term appearing in the last line 
of the $n=3$ matrix element, with no analogue in the Ricci flow, which  
contributes an additional term, without $\theta$, involving the operator 
${\cal D}_T^2 = -\partial / \partial t$ in the general power series expansion. 
The differences are becoming more visible in the structure of  
higher order terms, reflecting the more complicated nature of the commutation relations,  
which now involve a system of six rather than three basic bosonic generators 
of the algebra that all mix among themselves.    
 
It is relatively easy to perform the path-ordered integrations and 
work out the form of the power series expansion for 
axially symmetric deformations that depend only on the combination $Y=z+ \bar{z}$
for which the free field has the simpler looking form $\Phi_0(Y; t) = cY + d(t)$,  
as for the Ricci flow. For constant values of the parameter $c$, it turns 
the element $<T|M_+^{-1} M_- |T>$ can be computed as series expansion in 
powers of ${\rm exp} \Phi_0$ and the Toda superfield ${\cal F}$ assumes the 
following form     
\ba
{\cal F} & = & \Phi_0 + {1 \over (1! ~c)^2} {\cal D}_T e^{\Phi_0} + 
{1 \over (2! ~ c^2)^2} {\cal D}_T \left(e^{\Phi_0} {\cal D}_T e^{\Phi_0} \right) 
+ {1 \over (3! ~ c^3)^2} {\cal D}_T \left(e^{2 \Phi_0} {\cal D}_T^2 e^{\Phi_0} 
\right) \nonumber\\
& & + {1 \over (4! ~ c^4)^2} {\cal D}_T \left(e^{3 \Phi_0} {\cal D}_T^3 e^{\Phi_0} 
+ 11 e^{2\Phi_0} ({\cal D}_T e^{\Phi_0}) ({\cal D}_T^2 e^{\Phi_0}) \right) 
\nonumber\\
& & + {1 \over (5! ~ c^5)^2} {\cal D}_T \left( e^{4\Phi_0} {\cal D}_T^4 e^{\Phi_0} 
+ 29 e^{3\Phi_0} ({\cal D}_T^2 e^{\Phi_0})^2 \right) + \cdots    
\ea
by taking into account the contribution of the matrix 
elements \eqn{matele} up to order $n=5$. It is instructive to compare the terms with 
those appearing in the power series expansion \eqn{fitexp} 
for axially symmetric deformations 
of the Ricci flow. To each order we see that the terms transform from one into 
the other   
by simply changing $\partial / \partial t$ into ${\cal D}_T$. Polynomial terms of 
the form $(\partial_t {\rm exp} \Phi_0)^2$, and alike, are missing from the 
Calabi flow, since they will correspond to $({\cal D}_T {\rm exp} \Phi_0)^2$ 
and vanish for ordinary bosonic fields $\Phi_0$, due to $\theta^2 =0$. As for the 
remaining terms, there are some differences in their relative coefficients 
that appear to order $n \geq 4$ in the expansion and reflect the differences of 
the two algebras, as expected on general grounds.    
 
Substituting the result into the general expression \eqn{carim} 
with $\Psi_0 = 0$ and comparing the 
coefficients of the even and odd terms, the expansion of $\Phi (Y;t)$ 
in terms of free fields reads  
\be
\Phi = \Phi_0 - {1 \over (2! ~c^2)^2} e^{\Phi_0} \partial_t e^{\Phi_0} + 
{e^{2 \Phi_0} \over (4! ~c^4)^2} \left(e^{\Phi_0} \partial_t^2 e^{\Phi_0} + 
11 (\partial_t e^{\Phi_0})^2 \right) + \cdots ~,   
\ee
whereas the expansion for $\Psi(Y;t)$ reads, up to fifth order, 
\ba
\Psi & = & e^{-\Phi} \partial_Y^2 \Phi = 
-{1 \over (1! ~c)^2} \partial_t e^{\Phi_0} + {e^{\Phi_0} \over 
(3! ~c^3)^2} \left(e^{\Phi_0} \partial_t^2 e^{\Phi_0} + 
2(\partial_t e^{\Phi_0})^2 \right) - \nonumber\\
& & - {e^{2\Phi_0} \over (5! ~c^5)^2} \left(e^{2\Phi_0} \partial_t^3 e^{\Phi_0} 
+ 62 e^{\Phi_0} (\partial_t e^{\Phi_0})(\partial_t^2 e^{\Phi_0}) 
+ 87 (\partial_t e^{\Phi_0})^3 \right) + \cdots ~.   
\ea
The validity of these expressions can also be verified directly, order by order, by 
substituting into the system \eqn{altsyst} that describes the Calabi flow equation. 
However, unlike the Ricci flow, there are no simple mini-superspace 
solutions that describe axially symmetric deformations of the round sphere by the Calabi 
flow in closed form and which resemble the sausage model. 
It remains an interesting open question to find such models,  
provided that there are consistent mini-superspace reductions of the Calabi flow, and 
compare the exact solutions with the formal power series expansion above. Similar 
remarks also apply to solutions with non-trivial free field component $\Psi_0$ 
that are examined next.  

{\bf (ii) Solutions with $\Psi_0 \neq 0$}:    
In this case, the choice of fields should be 
compatible with the equations  
\be
e^{\Phi_0} \Psi_0 = \partial \bar{\partial} \Phi_0 ~, ~~~~~ 
\partial \bar{\partial} \Psi_0 = - {\partial e^{\Phi_0} \over \partial t}   
\ee
when applied to the most general 
free field configuration $(\Phi_0, \Psi_0)$. Of course, both 
sides of these equations vanish, since $\Phi_0$ is a two-dimensional free field that 
approximates well $\Phi$ only in a region of space where 
${\rm exp} \Phi_0 \simeq 0$, thus leaving $\Psi_0$ undetermined. 
Such asymptotic behavior is commonly 
assumed in all theories of Toda type in order for the formal power series expansion of
their solutions to be meaningful, if also convergent. 
A simple choice is to consider $\Psi_0$ independent of the coordinates $(z, \bar{z})$, as 
it will provide free field expansion for the configurations of Calabi flow around 
constant curvature spheres, with $R= -\Psi_0$, and in the vicinity of those points 
where $\Phi_0 \rightarrow -\infty$. The choice $\Psi_0 = 0$ that has been considered 
so far corresponds to such spheres, but with infinite radius, whereas in general one
may assume that $\Psi_0$ is non-zero, but it can also depend on time, $\Psi_0(t)$. 
Even more generally, $\Psi_0$ 
can be an arbitrary family of two-dimensional free fields that 
are independent of the choice $\Phi_0$. The analysis below is quite general, 
without assuming any restrictions on the form of $\Psi_0$, apart from the case of 
axially symmetric solutions that are only considered for illustrative purposes. 
Other interesting possibilities also arise by specializing to particular classes 
of free field configurations, as will be seen later.  
            
The calculations are considerably more involved now for evaluating the expectation   
value $<T|M_+^{-1}(z) 
M_-(\bar{z})|T>$ term by term in the power series expansion. Already, the first non-trivial 
term arising at order $n=1$ takes the form 
\ba
& & \int^{z} dz_1^{\prime} \int^{\bar{z}} d\bar{z}_1^{\prime} 
\int dt_1 \int dt_1^{\prime}  
{\rm exp} f(z_1^{\prime}; t_1) {\rm exp} \bar{f}(\bar{z}_1^{\prime}; t_1^{\prime}) 
\times \nonumber\\  
& & ~~~~~~~~~~ \times <T|[X_1^+(t_1) + \psi (z_1^{\prime}; t_1) X_0^+ (t_1)] 
[X_1^-(t_1^{\prime}) + \bar{\psi} (\bar{z}_1^{\prime}; t_1^{\prime}) 
X_0^- (t_1^{\prime})] |T>  
\ea
as can be readily seen using the path-ordered 
exponentials \eqn{mgefpoes} for $M_{\pm}$. For $\Psi_0 \neq 0$, there is 
an additional contribution coming  
from the cross terms in the expectation value, which were not present before.
For higher values of $n$ there are 
many more new terms that contribute to the final answer, making the process 
tedious. 

As before, it is also convenient here to restrict attention to axially symmetric 
configurations that depend only on the variable $Y = z+\bar{z}$ and assume that 
the two independent free fields are parametrized as follows,   
\be
\Phi_0 = c (z + \bar{z}) + d(t) ~, ~~~~~ \Psi_0 = a (z+ \bar{z}) + b(t) ~,  
\ee
where $a$ and $c$ are arbitrary constants whereas $b(t)$ and $d(t)$ depend generically 
on time. The path-ordered integrals can be evaluated easily in this case,  
but the resulting expressions 
are more lengthy than the simpler case with $\Psi_0 = 0$. 
It is quite instructive, 
nevertheless, to present the contributions to order 
$n=1$ and $2$,  
\ba
& & <T|M_+^{-1} M_- |T> = 1 - {1 \over (1! ~c)^2} e^{\Phi_0} \left( 1 + \theta \left( 
\Psi_0 - {2a \over c} \right) \right) + {1 \over (2! ~c^2)^2} 
[ \theta e^{\Phi_0} \partial_t e^{\Phi_0} \nonumber \\
& & ~~~~~~~ - \left(\Psi_0 -  
{3a \over c} -2 \right) e^{2 \Phi_0}  
- \theta \left(\Psi_0^2 - 4\left(1 + {a \over c}\right) \Psi_0 + {8a \over c} 
+ {7a^2 \over 2c^2} \right) e^{2 \Phi_0} ] ~,  
\ea
whereas higher terms are of order ${\rm exp}(3\Phi_0)$ or more.

Then, plugging into the general expression \eqn{carim} and comparing the even and 
odd parts of the equation, it follows to this oder that   
\be
\Phi = \Phi_0 + {1 \over (1! ~c)^2} \left(\Psi_0 - {2a \over c} \right) e^{\Phi_0} 
- {1 \over (2! ~ c^2)^2} \left( e^{\Phi_0} \partial_t e^{\Phi_0} 
- \left(\Psi_0^2 - {4a \over c} \Psi_0 + {7a^2 \over 2c^2} \right) 
e^{2\Phi_0} \right) 
\ee
and 
\be
\Psi = \Psi_0 - {1 \over (1! ~ c)^2} \partial_t e^{\Phi_0} - {1 \over (2! ~ c^2)^2} 
\partial_t \left(\left(\Psi_0 - {3a \over c} \right) e^{2\Phi_0} \right) .  
\ee
It can be verified, as independent check, that these expansions are consistent, order 
by order in powers of ${\rm exp} \Phi_0$, with the dimensionally reduced equations 
$\partial_Y^2 \Phi = \Psi {\rm exp} \Phi$ and $\partial \Psi = 
- \partial_t {\rm exp} \Phi$ 
accounting for axially symmetric deformations along the Calabi flow. Higher order 
corrections can also be included, if required. 

Finally, note in passing that there are special configurations in which both 
free fields $\Phi_0$ and $\Psi_0$ do not depend on time, $t$, and which 
can be used to provide 
power series solutions of the simpler equation $\Delta \Delta \Phi = 0$; 
equivalently they can be thought as describing B\"acklund transformations of 
the non-linear equation $\partial \bar{\partial} \Phi = \Psi_0 {\rm exp} \Phi$, 
which looks like a generalized Liouville theory with space dependent coupling 
$\Psi_0$. Our method is also capable to provide solutions of 
this equation in terms of an 
arbitrary free field $\Phi_0$ for any given field  
$\Psi_0$. We will say more about this time independent class of 
configurations, and their formal power series solutions, 
while discussing the so called type III Robinson-Trautman metrics in four 
dimensional general relativity.

\section{Robinson-Trautman space-times}
\setcounter{equation}{0}

In this section we discuss the embedding of the two-dimensional Calabi flow
into four-dimensional vacuum Einstein equations and review the main  
characteristics of the space-time solutions that are known up to this  
day. This material is included to make the presentation self-contained, as  
it allows to put our results in the physical context of spherical gravitational 
waves.      

\subsection{General structure of the metrics}

Recall that the most general gravitational vacuum solution in four space-time 
dimensions which admits a geodesic, shear-free, twist-free but 
diverging null congruence is given by the general class of 
Robinson-Trautman metrics, \cite{robi}  
\be
ds^2 = 2r^2 e^{\Phi} dz d\bar{z}  - 2 dt dr - H dt^2 ~, 
\label{muguy} 
\ee
where $H(z, \bar{z}, r, t)$ has the special form 
\be 
H =  r \partial_t \Phi - \Delta \Phi - {2 m(t) \over r} \label{frofac}  
\ee
expressed in terms of a single function $\Phi(z, \bar{z}; t)$, 
and $m(t)$ is a one-parameter family of 
parameters that in some cases represent the physical mass of the system. 
The affine variable $r$ varies along the rays of the repeated null eigenvector 
and $t$ is a retarded time coordinate. Closed surfaces with constant $r$ and $t$ represent 
distorted two-dimensional spheres, and, therefore, $t$-dependent solutions 
are thought to describe spherical gravitational radiation; of course, these are not 
exactly spherical gravitational waves 
in vacuum, since spherical symmetry implies that the 
above metric is static. It is more appropriate to think of 
the Robinson-Trautman metrics as
describing gravitational radiation outside some bounded region. 

There is a close analogy between Lienard-Wiechert fields of Maxwell equations and 
Robinson-Trautman metrics of Einstein equations in that they both admit a principal
null vector field which is geodesic, shear and twist free with non-vanishing 
divergence, thus giving a more intuitive meaning to the general ansatz that was made  
above, \cite{tedy}. 
In the gravitational case, the field equations imply that all vacuum solutions 
in this class satisfy the following fourth-order differential equation for 
$\Phi(z, \bar{z}; t)$,    
\be
\Delta \Delta \Phi + 3m \partial_t \Phi + 2 \partial_t m(t) = 0 ~, \label{rtrau}  
\ee
where $\Delta = {\rm exp}(-\Phi) \partial \bar{\partial}$ as before. 
To understand its equivalence to the Calabi flow in two dimensions, \eqn{cfe}, 
it suffices to relabel the null hypersurfaces of the Robinson-Trautman metrics 
and set $m$ equal to constant whenever $m(t) \neq 0$. The physically interesting 
case corresponds to positive values of $m$, in which case $t$ runs forward to infinity,    
but one may also consider the opposite case as well. Then, in the gauge with 
constant $m$, the Robinson-Trautman equation becomes  
\be
\Delta \Delta \Phi + 3m \partial_t \Phi = 0 \label{rtrau6}  
\ee
and it is identified with the Calabi flow by    
simple rescaling of the retarded time, so that $3m=1$; as such it provides an exact 
embedding of the two-dimensional Calabi flow into four-dimensional gravity, as it 
was first noted in reference \cite{tod}.   

Finally, we note for completeness that there are also interesting generalizations 
of the present framework to Robinson-Trautman space-times with cosmological  
constant $\Lambda$, which are obtained by modifying the function \eqn{frofac} 
appearing in the metric to $H_{\Lambda} = H - \Lambda r^2/3$. Vacuum Einstein equations
with cosmological constant reduce again to the same differential equation, 
as given in   
\eqn{rtrau}, which turns out to be independent of $\Lambda$. For an overview of the 
subject, see, for instance, \cite{kramer}.  

\subsection{Petrov classification of metrics}
   
Next, following \cite{kramer}, we review the Petrov classification of Robinson-Trautman vacuum 
solutions depending on the algebraic relations among the surviving components of the Weyl 
curvature tensor that are typically represented by the quantities
\ba
\Psi_2 & = & - {m \over r^3} ~, ~~~~~ \Psi_3 = {1 \over 2r^2} e^{-\Phi / 2} 
\bar{\partial} (\Delta 
\Phi) ~, \nonumber\\
\Psi_4 & = & {1 \over 2r^2} \bar{\partial} \left(e^{- \Phi} 
\bar{\partial} (r \partial_t \Phi - \Delta \Phi)
\right)   
\ea
and their complex conjugate expressions. There are four different algebraic types of  
space-times according to the list 
\ba
& & {\rm Type ~ N}: ~~~~ \Psi_2 = \Psi_3 = 0 ~, \nonumber\\
& & {\rm Type ~ III}: ~~~ \Psi_2 = 0 ~, ~~~ \Psi_3 \neq 0 ~, \\
& & {\rm Type ~ D}: ~~~~  3 \Psi_2 \Psi_4 = 2 {\Psi_3}^2 ~, ~~~ \Psi_2 \neq  0 ~, \nonumber\\
& & {\rm Type ~ II}: ~~~~ 3 \Psi_2 \Psi_4 \neq 2 {\Psi_3}^2 ~, ~~~ 
\Psi_2 \neq 0 ~. \nonumber   
\ea
The mass parameter $m$, which is taken constant, vanishes for the first two types and as a 
result the Robinson-Trautman equation reduces to 
\be
\Delta \Delta \Phi = 0 ~,  
\ee
with $t$ appearing as spectator without apparent role in dynamics. The last two types  
are characterized by the condition $m \neq 0$ and they correspond to 
genuine trajectories of the Calabi flow as given by equation \eqn{rtrau6}. More precisely, 
we have the following:  

{\em Type N solutions} satisfy additional conditions, by their definition, which 
give rise to the special equation 
\be
\partial \bar{\partial} \Phi = K(t) e^{\Phi} ~,  
\ee
using an arbitrary integration function of the variable $t$ alone, $K(t)$, 
that plays the role of 
coupling constant in the resulting two-dimensional Liouville equation. In this case, 
the $t$-dependence of space-times is not in general 
compatible with the Calabi flow, since $m=0$, unless $K(t)$ is constant.    
We only note here that flat space-time arises as special 
solution of this type, whereas non-flat type $N$ solutions typically 
have lines of singularities in the
three-dimensional $(r, z, \bar{z})$ space and they are also of limited interest 
in general relativity. 

{\em Type III solutions} satisfy the equation $\Delta \Delta \Phi = 0$, as before,  
with the condition  
\be
\bar{\partial} (\Delta \Phi) \neq 0 ~,  
\ee
and its complex conjugate, by their definition. Therefore, space-times of this type  
correspond to solutions of the differential equation 
\be
\Delta \Phi = f(z; t) + \bar{f} (\bar{z}; t) ~, \label{be3}  
\ee
where $f$ and $\bar{f}$ are complex conjugate functions of $z$ and 
$\bar{z}$, respectively. By reparametrization of the complex coordinates one may 
choose without lose of generality functions $f(z) = z$ and $\bar{f}(\bar{z}) = 
\bar{z}$ so that their characteristic equation becomes $\Delta \Phi = z+\bar{z}$. 
A particular simple solution is,   
\cite{robi},  
\be
e^{\Phi} = {3  \over (z + \bar{z})^3} ~. \label{spesol}  
\ee
More general static solutions of type $III$ are hard to construct beyond the ones that can
be obtained from \eqn{spesol} by arbitrary reparametrization 
of the coordinates $(z, \bar{z})$, whereas 
any dependence on $t$ is dropped out for all practical purposes.
   
{\em Type D solutions} are fully known and they include the Schwartzchild metric 
with mass parameter $m$ as static example. In this class, there are 
additional relations 
\be
\bar{\partial} \left( e^{-\Phi} \bar{\partial}  
(\Delta \Phi) \right) = 0 ~, ~~~~  
e^{-\Phi} \left(\bar{\partial} (\Delta \Phi) \right)^2 + 
3m \bar{\partial} \left( e^{-\Phi} 
\bar{\partial} (\partial_t \Phi) \right) = 0 ~, 
\ee
and their complex conjugate, which are obtained from the condition 
$3\Psi_2 \Psi_4 = 2 {\Psi_3}^2$ to different orders in the radial 
variable $r$. Integration of the first equation yields 
${\rm exp}(-\Phi) \bar{\partial} (\Delta \Phi) = f(z; t)$, where $f(z)$ 
is an arbitrary analytic function. Then, the Robinson-Trautman equation 
\eqn{rtrau6} simplifies to 
\be
\partial \left( f e^{\Phi} \right) +   
3m \partial_t e^{\Phi} = 0~,  
\ee
plus its complex conjugate relation. 

For $f = 0$ the corresponding metric 
is $t$-independent and takes the form
\be
ds^2 = {2r^2 \over \left(1 + {K \over 2} z \bar{z} \right)^2} dz d\bar{z} - 2dt dr - 
\left(K - {2m \over r} \right) dt^2  
\label{eddfilke}
\ee
with constant $K$ that can be normalized to $0$ or $\pm 1$. For $K=1$ it coincides with 
the Schwartzschild metric in Eddington-Finkelstein frame, setting   
\be
t = t_s - r - 2m {\rm log}(r-2m) 
\ee  
in terms of the usual time coordinate $t_s$ of the Schwartzschild solution.  
When $f$ is not zero, one can perform coordinate transformation to make it 
constant, say $f= -3m$, in which case the 
function $\Phi$ depends only on the combination
$z + \bar{z} + t$. Then, the geometry can be brought into the static 
$C$-metric form by further change of the 
coordinates (see, for instance, \cite{kramer}). 
A notable example of this kind corresponds to the choice, \cite{collin},       
\be
e^{\Phi} = {A \over (z + \bar{z} + t)^{3/2}} ~,   
\ee
having $4mA^2 = 1$.  

{\em Type II solutions} provide the last but most general class of space-times that are 
directly related to time dependent solutions of the Calabi flow.  
They admit arbitrary initial data $\Phi (z, \bar{z}; t_0)$ at some 
given time $t=t_0$, but unfortunately there are hardly any   
explicit solutions known to this day, apart from those obtained from 
type III configurations by allowing $m \neq 0$, \cite{robi}, as, for instance, the
particular solution \eqn{spesol}. 
Solution generating techniques have been subsequently employed toward the construction 
of more general type $II$ space-times, but the results are rather 
inconclusive in the literature; see, however, the recent work \cite{natorf} on 
type II metrics which provides some explicit solutions that are not asymptotically 
flat.
The main interest in type $II$ solutions stems from the fact that both gravitational 
radiation and black hole formation can be studied simultaneously in exact terms in 
the vacuum. 
Thus, it is an outstanding problem in general relativity to explore the general 
structure of type II solutions; here come our results to make a decisive step in 
this direction. 
 
\subsection{Further physical considerations}

It is clear from the discussion above that type III and type II Robinson-Trautman 
metrics provide the most general classes of gravitational metrics \eqn{muguy} 
in four dimensional space-times  
with parameter $m$ zero and non-zero, respectively, whereas type N and type D 
metrics follow from them as special limiting cases.  

Type II metrics provide a class of radiative space-times which tend  
to the Schwarzschild solution in the asymptotic future. Actually, 
approximate solutions of this kind were first examined in the 
literature on general relativity, \cite{newman},  
by perturbing the Schwarzschild metric so that the 2-spheres $M$ defined at constant 
$r$ and $t$ represent distorted configurations in the multi-pole expansion of 
gravitational radiation. In particular, using  
the linearized approximation, as outlined in section 2.5, 
small perturbations associated to spherical harmonics
with $l \geq 2$ were found to  
decay exponential fast by inducing quadrapole and higher order pole radiation,  
until the configuration eventually settles back to the round sphere as
$t \rightarrow \infty$. It was subsequently found that all dynamical type $II$ solutions 
evolving toward the Schwarzschild solution exhibit Lyapunov stability using the 
quadratic curvature functional $S(g)$, given by equation \eqn{quacurfu} for   
$M = S^2$, \cite{lukacs}. 
As was also shown in section 2, $S(g)$ varies monotonically with 
time and, in this case, the critical point where it assumes its
minimal value is the round sphere that corresponds to the Schwarzschild 
space-time in Eddington-Filkenstein frame \eqn{eddfilke}. In this case, the notion 
of extremal K\"ahler metrics that was introduced on general grounds by Calabi 
reduces to the constant curvature metric on $S^2$ that characterizes the end 
point of the dynamical evolution. Finally, the (semi)-global existence
and convergence of solutions with smooth initial data has been established in all
generality, following arguments similar to the long time existence and convergence
of solutions to the Ricci flow, thus establishing the universal limiting character
of all asymptotically flat type $II$ metrics, \cite{chrus} (but see also 
\cite{tod, schmidt} for important earlier work on the subject).           

The global structure of Robinson-Trautman space-times has been carefully analyzed
and extensions across the null hypersurface at $t = \infty$ have been considered  
in analogy to the Kruskal extension of the $r>2m$ Schwarzschild space-time.  
Two such solutions can be glued together along their Schwarzschild-like event horizon
to form space-times that contain both a black and a white hole, as in the ordinary 
static case. However, the extension through the horizon is not smooth, in general, 
which implies that an observer living in this space can determine by local measurements
whether or not he has crossed the horizon. These results were obtained by a 
detailed asymptotic analysis of the long time behavior of solutions to the  
Robinson-Trautman-Calabi equation, which upon transformation to Kruskal type
coordinates were found to exhibit terms in the power series expansion with logarithmic 
coefficients, \cite{piotro}. 
When the cosmological constant is not zero, the analysis of the 
solutions and their long time behavior essentially remains unaltered. However, the     
value of $\Lambda$ enters in the space-time interpretation of the solutions and their 
long time asymptotic convergence to the Schwarzschild solution with a cosmological 
constant, e.g., the Schwarzschild-de Sitter metric when $\Lambda >0$. An interesting 
feature of such generalization is that the continuation of the metric across the 
horizon can be made with a higher degree of smoothness when $\Lambda >0$, 
\cite{prague}, as 
compared to the more standard case with vanishing cosmological constant.    

Previous attempts in the literature to integrate the Robinson-Trautman 
equation for type II metrics include the application of prolongation 
methods, as in reference \cite{glass}. However, their results remain 
rather preliminary and inconclusive as they felt short of expectation 
compared to other two-dimensional integrable systems that can be more  
easily treated by such tools. There are also some works on series solutions 
of type II metrics, \cite{vand}, but a systematic understanding of their 
origin seems to be lacking. Our main contribution to this subject, 
according to the previous sections,  
is the ability to cast the class of all type II metrics into zero 
curvature form on the two-dimensional space $(z, \bar{z})$ and construct   
systematically formal solutions based on free fields, via group theoretical 
methods. The only drawback of our construction is the inability to characterize 
the global structure of the resulting space-times, and their possible 
singularities, in terms of the free field data used in the B\"acklund 
transformations. Also, simple but explicit solutions of the equations 
are still unknown, as they may never exist simple models of spherical 
gravitational radiation in vacuum that only depend on finite number 
of moduli.   

Type III metrics, on the other hand, require separate investigation as 
they do not correspond to genuine dynamical solutions of the Calabi flow.
True, they can be promoted to type II solutions of a very special type by 
allowing $m \neq 0$, but they also remain largely out of reach apart from 
specific solutions, such 
as \eqn{spesol}, and their descendants by coordinate transformations. 
The zero curvature formulation of type III metrics will be considered 
separately in the next section, where it is also explained how to 
extend the group theoretical methods of Toda theory in order to construct 
formal solutions in terms of free fields, as in the fully dynamical 
situation. We only note here that type III metrics are characterized by the 
non-vanishing component of the Weyl curvature tensor $\Psi_3$, which 
in turn implies that $\partial(\Delta \Phi)$ and $\bar{\partial}(\Delta \Phi)$ 
differ from zero. Thus, type III metrics do not correspond to extremal 
K\"ahler metrics on $S^2$, using the terminology introduced by Calabi, 
\cite{calabi1, calabi2}. On $S^2$ the extremal metrics are also constant 
curvature metrics with constant $\Delta \Phi$.

\section{The algebraic description of type III metrics}
\setcounter{equation}{0}

A zero curvature formulation of the non-linear equation that    
characterizes type III Robinson-Trautman metrics, 
\be
\Delta \Delta \Phi = 0 
\label{type33}
\ee
was given in reference 
\cite{bakala} by taking suitable limit of the commutation relations of the 
algebra used for the Calabi flow. Here, we provide another zero curvature 
formulation of the same equation, which is not equivalent to that construction, 
but is more natural for applying the 
general framework of Toda field theories and their solution generating techniques 
as in the Calabi flow.       

First, notice that the equation \eqn{type33} can be written in superfield 
form as 
\be
\partial \bar{\partial} {\cal F} = {\partial \over \partial \theta} e^{{\cal F}}  
\label{asstodf}
\ee
using ${\cal F} = \Phi + \theta \Psi$ with bosonic components that do not  
not depend on time, i.e., $t$ will be spectator if added by hand. 
Comparing the even and odd parts 
of the equation it follows that 
\be
\partial \bar{\partial} \Phi = \Psi e^{\Phi} ~, ~~~~~ \partial \bar{\partial} 
\Psi = 0 ~, 
\label{asstod7}
\ee
which is equivalent to equation \eqn{type33} after eliminating the field $\Psi$.  
Thus, equation \eqn{type33} admits a formal Toda field theory description with 
Cartan operator $\partial / \partial \theta$.  

Next, to appreciate the idea behind the new formulation of this  
equation that will be adopted in the sequel, 
let us briefly recall the essential features  
of Toda field equations based on the generalized system of commutation relations  
\ba
& & [X^+({\cal F}) , ~ X^- ({\cal G})] = H(S({\cal F} {\cal G})) ~, ~~~~ 
[H({\cal F}) , ~ H({\cal G})] = 0 ~, \nonumber\\
& & [H({\cal F}) , ~ X^{\pm} ({\cal G})] = \pm X^{\pm} ((K {\cal F}) {\cal G}) ~.   
\ea
Lie algebras of this type provide a natural supercontinual extension of the  
more abstract class of continual Lie algebras \eqn{bosalg} 
for any given pair of linear super-operators $(K, S)$. Of course, as in the 
ordinary bosonic case considered in section 3, a supercontinual Lie algebra with 
characteristics $(K, S)$ is isomorphic 
to that with $(\tilde{K} = KS, ~ \tilde{S} = 1)$,  
provided that the corresponding Cartan generators are related to each other 
by $\tilde{H} ({\cal F}) = H (S{\cal F})$ when $S$ is invertible. Thus, one 
typically makes the canonical choice $(\tilde{K}, 1)$ without loss of generality.  

The associated 
system of Toda field equations follows from the zero curvature condition 
$[\partial + A_+ , ~ \bar{\partial} + A_-] = 0$ choosing the  gauge connections 
\be
A_+ = H({\cal G}) + X^+(1) ~, ~~~~~ A_- = X^- (e^{\cal F}) ~.  
\ee
The resulting equations read, in general,  
\be
\partial {\cal F} = K {\cal G} ~, ~~~~~ \bar{\partial} {\cal G} = S e^{{\cal F}} 
\label{intequa}
\ee
and therefore one obtains 
\be
\partial \bar{\partial} {\cal F} = KS e^{{\cal F}} = \tilde{K} e^{{\cal F}} 
\ee
after eliminating the field ${\cal G}$. Note that the final field equation only 
depends on the operator $\tilde{K}$ and as such it is inert to the choice of 
algebra $(K, S)$ or $(\tilde{K}, 1)$. 
However, the intermediate equations \eqn{intequa},  
which transform into one another when different choices are being made,  
will be equivalent only if $S$ is invertible operator. 
 
Thus, it does not make a difference 
whether one uses $K=1$ and $S = {\cal D}_T$  or the canonical choice 
$K = {\cal D}_T$ and $S=1$, as we did before, for the zero curvature formulation 
of the Calabi flow.    
The situation changes drastically when the map $\tilde{H} ({\cal F}) = 
H(S{\cal F})$ is degenerate, as in the case of nilpotent operators with 
$S^2 = 0$, for the two algebras with operators $(K, S)$ and $(\tilde{K}, 1)$ are not 
anymore isomorphic. This particular situation is realized when 
$S= \partial / \partial \theta$ and the algebra with characteristics 
$(\partial / \partial \theta, 1)$ is not isomorphic to that with 
$(1, \partial / \partial \theta)$. Although the associated Toda field equation, 
which in this case reads as \eqn{asstodf}, does not know the difference,   
the intermediate equations \eqn{intequa} are actually sensitive on the choice. 
Hence, the natural question arises which is the most recommended choice of 
algebra in the present case.  

Let us choose $K=1$ and $S=\partial / \partial \theta$ and examine the detailed 
form of the equations \eqn{intequa}, which in terms of components 
${\cal F} = \Phi + \theta \Psi$ and ${\cal G} = g + \theta f$ become, respectively, 
\ba
& & \partial \Phi = g ~, ~~~~~ \partial \Psi = f ~, \nonumber\\
& & \bar{\partial} g = \Psi e^{\Phi} ~, ~~~~~ \bar{\partial} f = 0 ~.   
\ea
Then, by eliminating $f$ and $g$ one arrives at the system \eqn{asstod7} for the 
functions $\Phi$ and $\Psi$ that are required in all generality. 
On the other hand, the choice 
$K= \partial / \partial \theta$ and $S=1$ leads to a different system of intermediate
equations \eqn{intequa}, which in term of components read  
\ba
& & \partial \Phi = f ~, ~~~~~ \partial \Psi = 0 ~, \nonumber\\
& & \bar{\partial} g = e^{\Phi} ~, ~~~~~ \bar{\partial} f = \Psi e^{\Phi} ~.   
\ea
Eliminating $f$ and $g$, as before, we obtain the equations 
$\partial \bar{\partial} \Phi = \Psi e^{\Phi}$ and $\partial \Psi = 0$, 
which also lead to $\Delta \Delta \Phi = 0$ as required. However, the latter 
formulation is more restrictive than the first since $\partial \Psi = 0$, 
and its complex conjugate equation $\bar{\partial} \Psi = 0$,  
determine only a very special class of solutions to the second order 
equation $\partial \bar{\partial} \Phi = 0$ appearing in 
the system \eqn{asstod7}. 
Thus, generality requires making the choice $(1, \partial / \partial \theta)$ 
rather than $(\partial / \partial \theta, 1)$. The different choices 
yield equivalent descriptions only for constant curvature metrics in 
two dimensions, in which 
case the fourth order differential equation \eqn{type33} becomes equivalent 
to Liouville equation.      

With these explanations in mind, we choose the supercontinual Lie algebra 
with characteristics $K=1$ and $S=\partial / \partial \theta$, whose 
commutation relations take the  
following form in terms of components, using the decomposition of the 
generators $A=A_0 + \theta A_1$,    
\ba
& & [H_1(\varphi), ~ X_0^{\pm}(\psi)] = \pm X_0^{\pm} (\varphi \psi)  ~, 
~~~~ [X_1^{\pm}(\varphi), ~ X_0^{\mp}(\psi)] = \pm H_1 (\varphi \psi) ~, 
\nonumber\\
& & [H_0(\varphi), ~ X_1^{\pm}(\psi)] = \pm X_0^{\pm} (\varphi \psi) ~, 
~~~~ [H_1(\varphi), ~ X_1^{\pm}(\psi)] = 
\pm X_1^{\pm} (\varphi \psi) ~, 
\label{ktwo1}
\ea
whereas the rest are trivial,  
\ba
& & [X_0^+ (\varphi) , ~ X_0^- (\psi)] = 0 ~, ~~~~ 
[X_1^+ (\varphi), ~ X_1^{-} (\psi)] 
= 0 ~, \nonumber\\
& & [H_0 (\varphi) , ~ X_0^{\pm} (\psi)] = 0 ~, ~~~~~ 
[H_i (\varphi), ~ H_j (\psi)] = 0 ~. 
\label{ktwo2}
\ea
These commutation relations should be contrasted with the choice made in 
reference \cite{bakala}, which is also quite general, but it constitutes an 
inequivalent algebraic description of the same problem which will not be 
utilized in the present work. 

The formulation that is 
adopted here is more appropriate for integrating the equation 
$\Delta \Delta \Phi = 0$ by group theoretical methods, at least formally, as in 
ordinary Toda systems.    
The present description has the advantage that $t$ is completely decoupled from the 
commutation relations of the algebra, and, as a result, the smearing functions 
in the components of the generators, $A_0(\varphi)$ and $A_1(\varphi)$, can be 
completely dropped out. Put differently, the smearing superfields ${\cal F}$ can 
be considered as functions of $\theta$ only, without having any $t$ dependence, so 
that they correspond to super-numbers $\varphi_0 + \theta \varphi_1$. 
Also, the smeared generators can be simply defined by integration over 
the Grassmann variable 
$\theta$, as
\be
A({\cal F}) = \int A(\theta) {\cal F}(\theta) d\theta = \varphi_1 A_0 + 
\varphi_0 A_1 
\ee
and, as a result, the $\varphi$ and $\psi$ dependence in the above commutation 
relations is superfluous and factors out completely.
 
Likewise, the highest weight state that is formally introduced to integrate the 
associated Toda field equation \eqn{asstodf} assumes a simpler form, compared 
to the Calabi flow, as it 
can be taken independent of $t$. Thus, here, we consider a normalized 
vacuum state 
\be
|\theta> = |0>_0 + \theta |0>_1  
\ee
with defining relations, in smeared form,   
\be
X^+ ({\cal F}) |{\cal G}> = 0 ~, ~~~~ <{\cal G}| X^- ({\cal F}) = 0 ~, ~~~~ 
H({\cal F}) |{\cal G}> = |{\cal F} {\cal G}> ~,  
\ee
as before, where  
\be
|{\cal F}> = \int |\theta> {\cal F}(\theta) d\theta =  
\varphi_1 |0>_0 + \varphi_0 |0>_1 ~. 
\ee
In terms of components these relations read   
\ba
& & H_0 |0>_0 = 0 ~, ~~~~~~ H_0 |0>_1 = 
|0>_0 ~, \nonumber\\
& & H_1 |0>_i = |0>_i ~~~~ {\rm for} ~~ 
i=0, 1 ~,  
\ea
and 
\be
X_0^+ |0>_i = 0 = X_1^+ |0>_i ~~~~ {\rm for} ~~ i=0,1 
\ee
together with their Hermitian conjugate relations. 

Then, the group theoretical framework for Toda theories can be implemented     
in this case and the general solution of equation \eqn{asstodf} 
takes the form       
\be
{\cal F} (z, \bar{z}; \theta) = {\cal F}_0 (z, \bar{z}; \theta) -    
{\rm log} <\theta| 
M_+^{-1} (z) M_- (\bar{z})|\theta> ~,  
\label{kkttw} 
\ee
in accordance to the general expression \eqn{genex} applied to supercontinual 
Toda systems with $K=1$. 
${\cal F}_0 = \Phi_0 + \theta \Psi_0$, whose components are  
two-dimensional free fields
\be
\Phi_0 (z, \bar{z}) = f (z) + \bar{f} (\bar{z}) ~, ~~~~~
\Psi_0 (z, \bar{z}) = \psi (z) + \bar{\psi} (\bar{z}) ~.  
\ee
Finally, the operators $M_{\pm}$ are given by the path-ordered exponentials  
\ba
M_+ (z) 
& = & {\cal P} {\rm exp} \left(\int^{z} dz^{\prime} 
e^{f (z^{\prime})} [X_1^+ + 
\psi (z^{\prime}) X_0^+ ] \right) , \nonumber\\ 
M_- (\bar{z}) 
& = & {\cal P} {\rm exp} \left(\int^{\bar{z}} d\bar{z}^{\prime} 
e^{\bar{f} (\bar{z}^{\prime})} 
[X_1^-  + \bar{\psi} (\bar{z}^{\prime}) 
X_0^- ] \right)    
\ea   
that do not involve integration over the continuous time variable $t$. 

Actually, there are several simplifications that occur in this case. 
Note that the commutation relations \eqn{ktwo1} and \eqn{ktwo2} contain 
a subalgebra generated by $X_0^{\pm}$, $X_1^{\pm}$ and $H_1$, which is
sufficient to determine the structure of $<\theta | M_+^{-1} M_- | \theta>$. 
The power series expansion of the path-ordered exponentials 
$M_{\pm}$ contain strings of these generators only, and, as a result, 
the evaluation of the corresponding vacuum expectation values, 
which is carried out 
as usual by pushing $X^+$'s to the right and $X^-$'s to the left, can  
never involve $H_0$ and its action on the highest weight state $|\theta>$.    
Since $H_1|\theta> = |\theta>$, as opposed to $H_0|\theta> = \theta |0>_0$, 
it follows readily   
that the Grassmann variable $\theta$ never appears as coefficient in 
the resulting expressions, and, for all practical purposes, the vacuum 
expectation value can be restricted to the state $|0>_0$ alone. Thus, 
$\theta$ is also a superfluous variable in writing the general form of the 
solution, which simplifies to  
\be
\Phi (z, \bar{z}) = \Phi_0 (z, \bar{z}) -    
{\rm log} \left({}_0 \hspace{-1mm} <0| 
M_+^{-1} (z) M_- (\bar{z})|0>_0 \right) .   
\ee
At the same time, by comparing the odd terms of equation \eqn{kkttw}, 
we obtain to all orders the result  
\be
\Psi(z, \bar{z}) = \Psi_0 (z, \bar{z}) 
\ee
thanks to the absence of $\theta$-dependent terms in 
the expression ${}_0 \hspace{-1mm} <0|M_+^{-1} M_-|0>_0$.  

By construction, the power series expansion 
of a general field configuration $\Phi$ in terms of free fields 
$\Phi_0$ should coincide with the expansion that governs the Calabi 
flow when restricted to $t$-independent fields, since this limit is well 
defined within the general class of Toda systems. Thus, up to second order 
in powers of ${\rm exp} \Phi_0$, it follows that for axially symmetric 
configurations with $\Psi_0 = a(z+ \bar{z}) +b$ and $\Phi =
c(z+ \bar{z}) + d$, a broad class of solutions is    
\be
\Phi = \Phi_0 + {e^{\Phi_0} \over 
(1! ~ c)^2} \left(\Psi_0 - {2a \over c} \right) + {e^{2\Phi_0} \over 
(2! ~ c^2)^2} \left(\Psi_0^2 - {4a \over c} \Psi_0 + {7a^2 \over 2c^2} 
\right) + \cdots   
\ee
whereas $\Psi = \Psi_0$ to all orders in the expansion, as required. 
In the special case of constant curvature metrics, which 
correspond to $\Psi_0 = b$, the system \eqn{asstod7} reduces to the 
Liouville equation, 
\be
\partial \bar{\partial} \Phi = b e^{\Phi} 
\ee
and axially symmetric solutions of it assume the free field expansion 
\be
\Phi = \Phi_0 + {b \over (1! ~c)^2} e^{\Phi_0} + {b^2 \over (2! ~ c^2)^2} 
e^{2 \Phi_0} + \cdots 
\ee
according to the more general expression given above.      

Finally, it is interesting to identify the algebra whose local 
part is generated by $X_0^{\pm}$, $X_1^{\pm}$ and $H_1$ after dropping the 
superficial dependence on the smearing functions. Using the commutation 
relations \eqn{ktwo1} and \eqn{ktwo2}, and changing base to 
\ba
& & e_0 = {1 \over \sqrt{2}} (X_0^+ + X_1^+) ~, ~~~~~ 
e_1 = {1 \over \sqrt{2}} (X_0^+ - X_1^+) ~, \nonumber\\
& & f_0 = {1 \over \sqrt{2}} (X_0^- + X_1^-) ~, ~~~~~ 
f_1 = {1 \over \sqrt{2}} (-X_0^- + X_1^-) ~, 
\ea
it follows, setting also $H_1 = h$, that    
\be
[e_i ~ f_j] = \delta_{ij} h ~, ~~~~~ [h, ~ e_i] = e_i ~, ~~~~~ 
[h, ~ f_i] = - f_i ~.  
\label{ktwo3}
\ee
These are precisely the defining relations of Kac's $K_2$ algebra, 
\cite{victor}, which is 
the most elementary example of simple Lie algebra with infinite 
growth\footnote{The series of $K_n$ algebras is defined by the commutation 
relations \eqn{ktwo3} with $i$ and $j$ taking $n$ different values. In all 
cases there is only one Cartan generator $h$ and the growth of the 
algebras is infinite 
for $n \geq 2$. The case $n=1$ corresponds to the $sl(2)$ algebra 
and naturally it is not included in this series.}. 
Equivalently, it can be thought as arising from the algebra
\be
[e_i ~ f_j] = \delta_{ij} h_j ~, ~~~~~ [h_i, ~ e_j] = K_{ij} e_j ~, ~~~~~ 
[h_i, ~ f_j] = - K_{ij} f_j ~.  
\ee
with $i$ and $j$ taking values $0$ or $1$, using a degenerate 
$2 \times 2$ Cartan matrix
with all its elements equal to 1. In such case, $h_0 - h_1$ is a central element,  
as it commutes with all other generators $e_i$ and $f_i$, and when it is factored 
out yields the $K_2$ algebra.   

The  $K_2$ algebra was encountered in the literature before as the prolongation structure 
of the equation $\Delta \Delta \Phi = 0$ that characterizes type III Robinson-Trautman 
metrics, \cite{finley}. It is rewarding that it also arises here 
using a somewhat different route of investigation. Our general construction 
based on Toda theories also connects it nicely with 
the algebra that determines formal solutions of the Calabi flow. It should also be 
noted that there is yet another way to write the type III metrics in zero curvature 
form, following \cite{bakala}. The algebra used there is more complicated, as it 
depends on the additional parameter $t$ - thought as spectator - in a 
non-trivial way, and it only contains $K_2$ in its zero modes. The intertwining 
of these two different algebraic descriptions of the same equation deserves 
further attention as it may provide the means to construct non-trivial conservation 
laws. This problem is left open for future work.

\section{Hierarchy of higher order flows}
\setcounter{equation}{0}

Motivated by the general relation between the Ricci and Calabi flows, we introduce 
a hierarchy of higher order geometric flows by taking successive square roots 
of the time derivative operator. We will first describe the construction in 
two dimensions and outline an algorithm for casting the corresponding equations
into zero curvature form. Finally, we will discuss the generalization of the  
hierarchy of geometric evolution equations to K\"ahler manifolds of arbitrary dimension.  

\subsection{Higher order flows in two dimensions}

Let us assume that the conformal factor of the metric in two dimensions depends on 
infinite many time variables, 
$t_i$, as 
\be
\Phi = \Phi (z, \bar{z}; t_1, t_2, t_3, t_4, \cdots) ~, 
\ee
and that there is an associated hierarchy of intrinsic geometric flows, 
\be
{\partial \Phi \over \partial t_n} = \Psi_n (R) ~, \label{mashier}  
\ee
for appropriately chosen expressions of the scalar curvature $R = - \Delta \Phi$. 
The hierarchy is defined by imposing the following requirement 
for any two consecutive flows, 
\be
{\partial e^{\Phi} \over \partial t_{n+1}} = - 
{\partial^2 e^{\Phi} \over \partial t_n^2} ~. \label{taksqr}  
\ee

This, in turn, implies $\partial_{t_{n+1}} \Phi = - \partial_{t_n}^2 \Phi + 
(\partial_{t_n} \Phi)^2$,   
which suffices to determine recursively the form of 
curvature functionals $\Psi_n(R)$, starting from the Ricci flow 
that corresponds to the choice $\Psi_1 = -R$ for the time variable $t_1$. 
We have, in particular,  
\be
\Psi_{n+1} = - {\partial \Psi_n \over \partial t_n} - \Psi_n^2 ~, 
\ee
which upon iteration yields $\Psi_n$ in terms of $R$. Also note that due to equation 
\eqn{taksqr}, the compatibility condition of the proposed hierarchy 
reads $\partial_{t_{n+1}}(\Psi_n {\rm exp} \Phi) = \partial_{t_n} 
(\Psi_{n+1} {\rm exp} \Phi)$. Thus, in the appropriate sense, when the time derivatives 
act upon ${\rm exp} \Phi$ rather than $\Phi$, the hierarchy \eqn{mashier} provides 
a system of independent flows; 
in this respect, it could be compared to other systems  
of non-linear differential equations with increasing order, such as the KdV 
hierarchy.    

This procedure can be simply implemented as stated above. However, there is an alternative 
description in two dimensions that will allow to prove the integrability of these flows  
by uncovering the relevant algebraic structures, as was successfully done for the Calabi 
flow. In particular, let us start from $n$-th flow and make the substitution 
\be
{\partial \over \partial t_n} \rightarrow {\partial \over \partial \theta_n} 
- \theta_n {\partial \over \partial t_{n+1}} ~, ~~~~~ 
\Phi \rightarrow \Phi + \theta_n (\partial_{t_n} \Phi)  
\ee
with the aid of a Grassmann variable $\theta_n$, so that the $(n+1)$-flow is 
realized as super-evolution partner of the $n$-th flow. As a result, the order  
of the resulting parabolic equation is doubled, so that the $n$-th member in the
family of flows \eqn{mashier} has order $2^n$. The substitution made for $\Phi$ 
can be understood as requirement of its super-analyticity on the super-coordinate 
$T_n = (t_n, \theta_n)$ that extends $t_n$ in $R^{1|1}$ superspace; of course,  
$\partial_{t_n} \Phi = \Psi_n$, which in turn determines $\Psi_{n+1}$ 
by working out the form of the equation in components. Also, consistency of the scheme 
requires that the Grassmann variables $\theta_i$ introduced in each step are all 
independent, i.e., $\theta_i \theta_j + \theta_j \theta_i = 2 \delta_{ij}$. 
     
We have already seen in detail how this method yields the Calabi flow with respect
to the time variable $t_2$ (in the present notation). 
Further iteration yields the following eighth order 
differential equation, 
\be
{\partial \Phi \over \partial t_3} = 
\Delta [(\Delta \Phi)(\Delta \Delta \Phi) - 
\Delta \Delta \Delta \Phi]  
= \Delta \left(R\Delta R + \Delta \Delta R \right) ~, \label{eigthord}  
\ee
using the expression $\Psi_2 = - \Delta \Delta \Phi$ for the Calabi flow. The 
integrability of this equation will be established later by implementing its 
algorithmic construction to the level of supercontinual Lie algebras. 

The structure of the equations becomes considerably more involved for 
higher values of $n$, as their order increases exponentially. We only 
give here the next one, for illustration,    
\ba
{\partial \Phi \over \partial t_4} & = & 
\Delta [(\Delta \Phi) \left(\Delta (\Psi_3 \Delta \Phi) - 
\Delta \Delta \Psi_3 + 2 \Psi_3 \Delta \Delta \Phi \right) - (\Delta \Psi_3) 
(\Delta \Delta \Phi) \nonumber\\ 
& & - ~ \Psi_3 \Delta \Delta \Delta \Phi + \Delta \left(\Delta \Delta \Psi_3 - 
\Delta (\Psi_3 \Delta \Phi) - \Psi_3 \Delta \Delta \Phi \right)] ~,  
\ea
where $\Psi_3 = \Delta ((\Delta \Phi) (\Delta \Delta \Phi) - \Delta \Delta \Delta \Phi)$. 
Higher order equations can be worked out at will, but their structure cannot be 
easily described in closed form. Unfortunately, we have no systematic way to iterate 
the procedure and obtain $\Psi_n$ for all values of $n$. Likewise, the choice of 
supercontinual Lie algebras that enables to cast them into zero curvature will only be 
prescribed algorithmically.   

Finally, one may formally append in the beginning of this list the heavenly equation,
which is not parabolic but yields the Ricci flow by taking the square root of its 
time derivative operator.  
 
\subsection{Integrability of higher order flows}

In order to understand how the hierarchy of higher order flows can be brought 
into zero curvature form, let us first consider the next example in the list given by   
the eighth order equation \eqn{eigthord}. In this case, the formal substitution 
$\partial / \partial t_2$ by $\partial / \partial \theta_2 - 
\theta_2 \partial / \partial t_3$ supplemented by the change of field variable  
$\Phi (z, \bar{z}; t_2)$ to superfield ${\cal F}(z, \bar{z}; t_3, \theta_2)$ 
can be directly applied to the continual Lie algebra for the Calabi flow. 
Using its description in terms of components $H_i(\varphi)$ and $X_i^{\pm}(\varphi)$ 
with $i=0, 1$, as given in section 5, one obtains the following supercontinual 
Lie algebra with basic commutation relations      
\ba
& & [H_0({\cal F}), ~ X_0^{\pm}({\cal G})] = \pm X_0^{\pm} ({\cal F}{\cal G}) ~, 
~~~~ [X_1^{\pm}({\cal F}), ~ X_0^{\mp}({\cal G})] = \pm H_0 ({\cal F}{\cal G}) ~, 
\nonumber\\
& & [H_0({\cal F}), ~ X_1^{\pm}({\cal G})] = \pm X_1^{\pm} ({\cal F}{\cal G}) ~, 
~~~~ [X_1^+ ({\cal F}), ~ X_1^- ({\cal G})] = H_1 ({\cal F}{\cal G}) ~, 
\nonumber\\
& & [H_1({\cal F}), ~ X_1^{\pm}({\cal G})] = 
\mp X_0^{\pm} (({\cal D}_T {\cal F}) {\cal G}) ~, 
\ea
whereas the rest are trivial,  
\ba
& & [H_1({\cal F}), ~ X_0^{\pm} ({\cal G})] = 0 ~, ~~~~ 
[X_0^+ ({\cal F}), ~ X_0^{-} ({\cal G})] 
= 0 ~, \nonumber\\
& & [H_i ({\cal F}), ~ H_j ({\cal G})] = 0 ~. 
\ea
Here, to simplify notation, we drop the indices from the new time variable $t_3$ and 
the associated Grassmann variable $\theta_2$ and denote the corresponding 
super-evolution operator by ${\cal D}_T$, as before. 

Then, starting from the zero curvature formulation of the Calabi flow 
written in terms of 
components and promoting the smearing functions to superfields, so that the gauge 
connections are taking values in the new supercontinual Lie algebra given above, 
we immediately arrive to the zero curvature formulation of equation \eqn{eigthord}. 
In particular, after some simplification of the original expressions used for the 
Calabi flow, we are led to propose the following choice of gauge connections,
\ba
A_+ & = & H_0 (\partial {\cal F}) + H_1(\partial \omega) + 
X_1^+(1) ~, \nonumber\\
A_- & = & X_0^- (\partial \bar{\partial} {\cal F}) + X_1^- (e^{{\cal F}}) ~, 
\ea
where ${\cal F}$ and $\omega$ are now assumed to be $R^{1|1}$ mixed superfields 
with bosonic components 
that satisfy the relation 
\be
{\cal D}_T \omega = - e^{-{\cal F}} \partial \bar{\partial} {\cal F}.   
\ee

The zero curvature condition $F_{z \bar{z}} = 
[\partial + A_+, ~ \bar{\partial} + A_-] = 0$ 
becomes equivalent to the following equation for the superfield 
${\cal F} (z, \bar{z}; T)$   
\be
{\cal D}_T e^{{\cal F}}= - \partial \bar{\partial} \left( e^{-{\cal F}} \partial 
\bar{\partial} {\cal F} \right) , 
\ee
which when expanded in terms of components ${\cal F} = \Phi + \theta \Psi$ yields
the system
\be
\partial_t e^{\Phi} = \partial \bar{\partial} \left( \Delta \Psi - \Psi \Delta \Phi 
\right) , ~~~~~ \Psi = -\Delta \Delta \Phi ~.  
\ee
Eliminating $\Psi$ leads to the eighth order evolution equation \eqn{eigthord}, 
as advertised. 

One can work out the form of the gauge connections, as well as the 
form of the supercontinual Lie algebra, in components by expanding everything
in $R^{1|1}$ superspace. This will give rise to four bosonic components   
for each type of basic generators, which can be labeled as $H_{ij}(t)$ and 
$X_{ij}^{\pm}(t)$ with $i$ and $j$ taking values $0$ or $1$.   
The resulting expressions are quite lengthy and they are not included here. 
These generators, however, can be further promoted to elements of a supercontinual 
Lie algebra by changing $\partial / \partial t$ to ${\cal D}_T$ and using $R^{1|1}$ 
superfields as new smearing functions. This procedure can be repeated 
indefinitely, and, as a result, one obtains a recursive construction of an 
infinite hierarchy of continual Lie algebras that are associated to the hierarchy 
of higher order flows. Thus, any member of the family is cast into zero 
curvature form by the appropriate choice of algebra. 

In general, a flow of order $2^{n+1}$, which is associated to a 
deformation variable $t$, is 
naturally obtained from a continual Lie algebra whose basic elements can be  
labeled as follows,
\be
H_{i_1 i_2 \cdots i_n} (t) ~, ~~~~~ X_{i_1 i_2 \cdots i_n}^{\pm} (t) ~,  
\ee
with all indices taking values $0$ or $1$. There are $2^n$ such independent elements 
for each type of generators, which follow by iterating the decomposition of $R^{1|1}$ 
superfields used at each level. Alternatively, they can be regarded as the  
bosonic components (in appropriate base) 
of three basic generators $H(T)$ and $X^{\pm}(T)$, which 
are labeled by a supercontinuous index in $R^{1|n}$ superspace with coordinates 
$T=(t; \theta_1, \theta_2, \cdots , \theta_n)$, and likewise for their 
smearing functions. Thus, one may also consider the decomposition 
\be
A(T) = A_0(t) + \sum_{i=1}^n \theta_i A_i(t) + \sum_{i<j} \theta_i \theta_i 
A_{ij}(t) + \cdots + \theta_1 \theta_2 \cdots \theta_n A_{12 \cdots n}(t)  
\ee
applied each basic generator $H(T)$ or $X^{\pm}(T)$. The coefficients in the expansion 
are fully anti-symmetric in their indices, and, therefore, the number of independent 
components for each generator also sums up to $2^n$ in this parametrization. 

It will be interesting to see whether the resulting hierarchy of infinite 
dimensional algebras can be put in more compact form in terms of $H(T)$, 
$X^{\pm}(T)$ so that the standard group theoretical methods of Toda field equations 
can be directly applicable to them for appropriate choice of Cartan operator.  
These steps will not be carried out explicitly here 
for the higher order equations but are left for the future.

\section{Generalization to arbitrary dimensions} 
\setcounter{equation}{0}

The hierarchy of geometric flows can be generalized to all  
K\"ahler manifolds of arbitrary dimension. Indeed, given the   
formal relation between the Ricci and Calabi flows, as outlined in all generality 
in section 2, we are led to propose a hierarchy of higher order geometric flows of the form
\be
{\partial \over \partial t_n} g_{a \bar{b}} = \Omega_{a\bar{b}}^{(n)} (R)  
\ee
imposing the requirement $\partial_{t_n}^2 g_{a\bar{b}} = - \partial_{t_{n+1}}g_{a\bar{b}}$
for all time variables $t_n$. Everything that was said before generalizes quite naturally  
to all dimensions, apart from the underlying algebraic structures that are only useful 
for the zero curvature formulation of the flows on  
K\"ahler manifolds of complex dimension $1$. 
We already have $\Omega_{a\bar{b}}^{(1)} = -R_{a\bar{b}}$ 
and $\Omega_{a\bar{b}}^{(2)} = \partial_a \bar{\partial}_b R$, and therefore  
all higher order equations assume the general form
\be
{\partial \over \partial t_n} g_{a\bar{b}} = 
\partial_a \bar{\partial}_b V_n 
\ee
for appropriately chosen expressions $V_n(R)$. 

To illustrate the general structure of the equations, let us construct the next member 
of the hierarchy by computing the Calabi velocity of $R$, namely     
\be
{\partial \over \partial t_2}R = - \Delta \Delta R - R^{, a\bar{b}} R_{a \bar{b}} ~.  
\ee
It is convenient at this point to introduce the fourth order elliptic operator of 
Lichn\'erowicz, which is defined for each K\"ahler manifold as 
\be
{\cal L} \varphi = D_L \varphi - g^{a \bar{b}} R_{, a} \varphi_{, \bar{b}}.  
\ee
According to the definition of $D_L$ we have ${\cal L}R = \Delta \Delta R 
+ R^{, a\bar{b}} R_{a\bar{b}}$, and, therefore, the next flow takes the form 
\be
{\partial \over \partial t_3} g_{a \bar{b}} = 
\partial_a \bar{\partial}_b \left({\cal L}  
 R \right) ~.  
\ee
It is a parabolic equation with respect to the new time variable $t_3$ and provides   
an eight order non-linear differential equation for the components of the K\"ahler metric. 
Similarly, one may work out the form of higher order flows by iteration, but the final results 
are rather complicated. In general, the $n$-th flow is a parabolic equation of order 
$2^n$ with respect to the space coordinates.

In all cases, apart from the (unnormalized) Ricci flow, the deformations are 
volume preserving, i.e.,  
\be
{\partial \over \partial t_n} \int_M dV(g) =  
\int_M dV(g) \Delta V_n = 0  
\ee
for $n \geq 2$. They also  
preserve the K\"ahler class of the metrics as for the Ricci and Calabi flows.
In fact, for all these flows, the quadratic curvature functional $S(g)$ that 
was introduced in section 2 exhibits an absolute minimum value equal to 
$(\int_M R~ dV(g))^2 / \int_M dV(g)$ that is attained by constant curvature 
metrics, when they exist on the K\"ahler manifold $M$. Also, the extremal 
metrics on $M$ arise as local minima of $S(g)$ for all higher order flows 
as well. 
 
More precisely, using the form of the flows $\partial_{t_n} g_{a \bar{b}} = 
\partial_a \bar{\partial}_b V_n (R)$, we first compute the variation of 
$S(g)$, 
\ba
\partial_{t_n} S(g) & = & {\partial \over \partial t_n} \int_M R^2 [g] dV(g) = 
-2 \int_M R (D_L V_n) dV(g) \nonumber\\
& =& -2 (D_L V_n, ~ R) = -2 (V_n, ~ D_L R) ~.   
\ea
The result is obtained, as in section 2, by introducing the fourth order elliptic
operator $D_L$ and removing the total derivative terms that arise in the course 
of the calculation. The Euler-Lagrange equation $D_L R = 0$ describes the 
extremal values of $S(g)$ for all such flows, and the corresponding configurations 
are the extremal metrics on $M$, when they exist, 
satisfying the equivalent (but simpler) 
equation $L R = 0$, as for the Calabi flow. Actually, all these extrema are 
local minima because one finds that the Hessian form of $S(g)$ in the 
directions $t_n$ and $t_m$ is 
\be
{\partial^2 S(g) \over \partial t_n \partial t_m} = 2 (D_L \bar{D}_L V_n , ~ 
V_m)  
\ee
when evaluated at the critical points. The operator $D_L$ commutes with its 
complex conjugate $\bar{D}_L$ when a critical metric is being used and 
$D_L \bar{D}_L$ is a strongly elliptic eighth order operator that is 
self-adjoint and positive semi-definite. Then, for any given flow, the second 
variation is manifestly positive semi-definite,
\be
{\partial^2 S(g) \over \partial t_n^2} = 2 (V_n, ~ D_L \bar{D}_L V_n) \geq 0 ~, 
\label{hess} 
\ee
implying that all extrema are local minima, as required.      

It is fair to say that the preceding analysis appeared for the first time in the 
original works of Calabi, \cite{calabi1, calabi2}, 
who considered the general problem of deforming 
K\"ahler metrics on $M$ within a given cohomology class, 
\be
\tilde{g}_{a \bar{b}} = g_{a \bar{b}} + \partial_a \bar{\partial}_b u ~,  
\ee
using arbitrary $t$-dependent globally defined scalar functions $u(z, \bar{z})$.  
The hierarchy we have presented above corresponds to specific choices of $u$, as   
given by the potential functions $V_n(z, \bar{z}; t_n)$. The ordinary 
Calabi flow corresponds to the choice $V_2(z, \bar{z}; t_2) = R$ and has 
the special property that $S(g)$ decreases monotonically along it. All 
flows have critical points which are local (and possibly global) 
minima of $S(g)$ associated to extremal K\"ahler metrics. We see that the 
only directions of deformation of a critical metric in which the expression 
\eqn{hess} vanishes are those tangent to the orbit formed by all 
holomorphic transformations of $M$ preserving the K\"ahler class of the metric.   
These transformations act as gauge group for the variational problem of
minimizing the curvature functional $S(g)$ for they leave it invariant. 
Conversely, any flow that deforms the metric transversely to the gauge 
group orbit has strictly positive definite second derivative, as given by 
equation \eqn{hess}.   
 
Finally, note that there are no curvature functionals in our disposal, 
at least for the moment being, which decrease 
monotonically along the corresponding higher order flows. 
It is an interesting variational problem to 
construct such expressions and study the long time 
behavior of their trajectories, as for 
the ordinary Calabi flow. Further implications of the 
higher order geometric deformations to problems 
of K\"ahlerian geometry are left open for future work together 
with their possible physical applications.

\section{Conclusions and discussion}
\setcounter{equation}{0}

We have provided a unifying framework for formulating geometric evolution equations 
driven by the intrinsic curvature on K\"ahler manifolds, so that the metric deforms 
to canonical form within a given cohomology class. In two dimensions, in particular, 
these equations admit zero curvature description with gauge connections taking values 
in appropriately chosen infinite dimensional Lie algebras that incorporate the 
deformation variable $t$ into their system. Such uneven treatment of the coordinates 
$(z, \bar{z})$ and $t$ allows to establish an algorithm for their formal integration 
in terms of a one-parameter family of two dimensional free fields, as in Toda field 
equations. At the same time, the evolution of any given initial metric exhibits 
dissipative behavior in $t$, since curvature perturbations around the canonical 
metric tend to decay exponentially fast. However, there is no contradiction of terms 
with the notion integrability in two dimensions, as $t$ becomes internal index of the 
algebra of gauge transformations in the corresponding zero curvature conditions. 
Thus, it is possible to bring the equations into two-dimensional integrable form and
hide all dissipative behavior in $t$ in the internal space. 

The notion of supercontinual Lie algebras that was 
introduced here for the first time, has been particularly useful for establishing 
the zero curvature formulation of the two-dimensional Calabi flow by extending 
previously known results for the Ricci flow. This was implemented in practice by 
by extending $t$ to $R^{1|1}$ superspace with coordinates $(t, \theta)$ 
and promoting the time evolution 
operator to super-evolution. Further iteration of this procedure gave rise to an 
infinite hierarchy of higher order geometric flows, which exist on K\"ahler manifolds 
in all dimensions and they are integrable in two dimensions in the same sense that 
the Ricci and Calabi flows are. Actually, all these flows provide special examples
of a more general variational problem posed by Calabi for deforming the metrics 
within a given K\"ahler class. In that context, one considers 
the quadratic curvature functional for a given K\"ahler manifold $M$ of 
arbitrary dimension, whose local 
minima define the notion of extremal metrics on $M$, such as constant curvature 
metrics, provided that there are no 
obstructions for their existence. The general class of flows introduced by 
Calabi, $\partial_t g_{a \bar{b}} = \partial_a \bar{\partial}_b u$,   
exhibit common qualitative behavior in that they deform any initial  
geometric configuration toward the extremal metrics. The standard Calabi flow 
corresponds to the deformation driven by the  
scalar curvature, $u=R$, but there can be more arbitrary choices of the 
$t$-dependent scalar function $u$. The hierarchy we have constructed here selects 
some special functions $u$ in a recursive way, such as $R$, ${\cal L}R$, etc, 
and the corresponding higher order  
flows were found to admit zero curvature formulation in two dimensions. 
It will be interesting 
to examine whether more general Calabi deformations can admit similar 
description in two dimensions using appropriately chosen infinite dimensional 
Lie algebras depending on the form of $u (z, \bar{z}; t)$.       
    
The infinite dimensional structures that arise in the algebraic description of 
two-dimensional flows are quite novel, as they typically exhibit infinite growth. 
The prime example is the continual Lie algebra with Cartan operator 
$K=\partial / \partial t$ used for the Ricci flow, which grows exponentially fast 
beyond its local part. The supercontinual algebra with $K={\cal D}_T$ that was 
used for the Calabi flow is also known to have infinite growth, and the same  
behavior is expected from the infinite dimensional algebras associated to 
all higher order flows. Luckily, the zero curvature formulation of these flows 
involves gauge connections that take values in the local part of the algebra, 
and, hence, the complete structure of the algebra is not required for their Toda-like 
description and the formal construction of their general solution. However, it 
is natural to expect that the explicit construction of two-dimensional 
integrals requires better handle on the commutation relations of the 
algebras used in each case, beyond their local part, and the existence of invariant 
forms. It is conceivable that there will be no algebraic integrals of these 
equations, as in some other examples of exactly solvable systems whose zero 
curvature formulation is based on infinite dimensional gauge groups, 
like the long-studied Halphen equations, \cite{leon}. 
These issues remain out of reach at the moment, 
as more mathematical work is required 
to bring the theory of such algebras, and their representations, to higher 
level of understanding. More generally, the very notion of integrability for 
two-dimensional systems based on infinite dimensional gauge algebras calls for 
further work, as there are only some isolated examples in the literature 
so far; see, also, references \cite{takeo} and \cite {ablow} where other 
examples are analyzed and compared to more traditional systems whose integrability 
properties are compatible with the Painleve criterion.   
  
The main physical application of the present work lies within the theory of 
spherical gravitational waves in vacuum in four space-time dimensions, as 
given by the general class of radiative Robinson-Trautman metrics. 
The formal integration of the Robinson-Trautman (two-dimensional Calabi) equation
by group theoretical methods allows to express all type II solutions (in 
Petrov's classification) using two-dimensional  
free fields, as in all Toda field equations. However, there are hardly any 
explicit solutions known to this day that may depend on a finite number of moduli 
of the deformed two-dimensional sphere sitting inside the four-dimensional 
radiative metrics. If that were the case, some simple solutions of gravitational 
radiation in the exterior of bounded sources would have been possible to 
manufacture and study in exact terms. Put differently, there seems to be no 
reasonable mini-superspace description of the Robinson-Trautman-Calabi equation, 
unlike the simpler case of two-dimensional Ricci flows 
that admit consistent truncations and lead  
to interesting axially symmetric deformations of the round sphere, as 
the sausage model. Mini-superspace truncations of non-linear evolution 
equations, when they exist, are 
also tractable by more direct computational methods that do not rely on the 
Toda field theory interpretation of the equations and the 
free field parametrization of their general solution. Thus, it appears that 
exact models of gravitational radiation depend on an infinite number of 
parameters and there is only a formal power series expansion for them based on 
the group theoretical interpretation of the general solution. The class 
of type III Robinson-Trautman metrics was also shown to define an  
integrable two-dimensional system with gauge connections taking values  
in an infinite dimensional algebra identified with Kac's $K_2$ algebra 
of infinite growth.   

Another interesting problem for future work is the generalization of the 
present formalism to include geometric evolution equations on 
supermanifolds. 
So far, there has been no systematic work in this direction and it 
certainly deserves more attention. A seemingly related problem is the 
supersymmetric generalization of the Ricci and Calabi flows on 
K\"ahler manifolds. It will be interesting to examine whether in two 
dimensions, in particular, there are appropriate choices of continual 
and supercontinual Lie superalgebras that can accommodate such 
generalizations in the form of zero super-curvature conditions. A different  
but illustrative example is provided by the $N=1$ supersymmetric generalization 
of the heavenly equation, whose bosonic version is 
viewed as continual Toda system in two dimensions 
with gauge connections taking values in the local part of the $SU(\infty)$ 
algebra. In the supersymmetric extension of the heavenly equation, 
\cite{sorba}, one considers the Lie superalgebra 
$sl(N|N+1)$ in the large $N$ limit in which the system of simple roots is odd 
and corresponds to the super-principal embedding of $osp(1|2)$ in it. As 
a result, the Cartan operator of the corresponding continual 
superalgebra is $\partial / \partial t$, equal to the 
square root of the Cartan operator of $SU(\infty)$, and it happens to coincide 
with the Cartan operator of the continual Lie algebra used for the Ricci 
flow. This occurrence might be only the beginning of a systematic pattern for 
constructing supersymmetric extensions of the hierarchy of higher order 
geometric flows, together with their zero curvature 
formulation, by also turning ordinary derivatives into supersymmetric 
covariant derivatives in two dimensions. It remains to be seen how far 
this construction can be pushed and what is the systematics behind it.        

The use of infinite dimensional algebras that incorporate the 
time variable $t$ into their defining system of commutation relations 
might be of more general value for the reformulation of many dynamical 
problems other than the geometric evolution equations. If this possibility 
materializes it will certainly provide new ways to address difficult 
dynamical problems in terms of new classes of infinite dimensional 
Lie algebras and their representations. It may also fertilize the 
interplay between physics and mathematics in an area where Lie algebras 
of infinite growth become relevant and demand better understanding of 
the whole subject.   

Finally, it is quite interesting to consider the interplay between 
physics and mathematics for other classes of geometric evolution equations 
and develop new tools for their study. In this context, a 
unified picture of intrinsic as well as extrinsic flows could emerge.  

\vskip1cm
\centerline{\bf Acknowledgements}

This work was supported in part by the European Research and Training Network 
``Constituents, Fundamental Forces and Symmetries of the Universe" under contract number
MRTN-CT-2004-005104 and the INTAS program ``Strings, Branes and Higher Spin Fields" 
under contract number 03-51-6346. I thank  
the Theory Division at CERN for hospitality and financial support during my 
sabbatical leave in the academic year 2004-05, where the main body of the present 
work was carried out in excellent and stimulating environment.  

\vskip1cm
\centerline{\bf Dedication}

This paper is dedicated to the glowing memory of an exceptional man and a great 
scientist, Bryce S. DeWitt.

\end{document}